\documentclass[fleqn,usenatbib,usedcolumn]{mnras}
\usepackage{graphicx}	
\usepackage{amsmath}	
\usepackage{amssymb} 
\usepackage{times} 
\usepackage{aas_macros} 
\usepackage[dvipsnames,usenames]{color} 
\usepackage{longtable}
\usepackage{array}
\usepackage{soul}

\newcommand{\ACCnb}{in Paper~I}

\def\itcf{{\it cf.\ }}
\def\itie{{\it i.e.\ }}

\def\note #1]{\noindent{\bf #1]}}

\def\sgnk2{{\rm sgn\left(K^2\right)}}

\def\rmpsd{{\rm PSD}}

\title[Stochastic Signatures II]{From Solar-like to Mira stars: {a unifying} description of stellar pulsators in the presence of stochastic noise} 

\author[Cunha al.]{M. S. Cunha$^{1,3}$, P. P. Avelino$^{1,2,3}$, W. J. Chaplin$^{3,4}$\\
$^{1}$Instituto de Astrof\'\i sica e Ci\^encias do Espa\c co, Universidade do
Porto, CAUP, Rua das Estrelas, PT4150-762 Porto, Portugal\\
$^{2}$Departamento de Física e Astronomia, Faculdade de Ciências, Universidade do Porto, Rua do Campo Alegre 687, PT4169-007 Porto, Portugal\\
$^3$School of Physics and Astronomy, University of Birmingham, Birmingham, B15 2TT, United Kingdom\\
$^4$Stellar Astrophysics Centre (SAC), Department of Physics and Astronomy, Aarhus University, Ny Munkegade 120, DK-8000 Aarhus C, Denmark\\
}

\date{Accepted XXX. Received YYY; in original form ZZZ}

\pubyear{2019}

\begin{document}
	\label{firstpage}
	\pagerange{\pageref{firstpage}--\pageref{lastpage}}
	\maketitle
\begin{abstract}
{We discuss and characterise the power spectral density properties of a model aimed at describing pulsations in stars from the main-sequence to the asymptotic giant branch. We show that the predicted limit {of the} power spectral density for a pulsation mode in the presence of stochastic noise is always well approximated by a Lorentzian function. While in stars predominantly stochastically driven the width of the Lorentzian is defined by the mode lifetime, in stars where the driving is predominately coherent the width is defined by the amplitude of the stochastic perturbations. In stars where both drivings are comparable, the width is defined by both these parameters and is smaller than that expected from pure stochastic driving. We illustrate our model through numerical simulations and propose a well defined classification of stars into predominantly stochastic (solar-like) and predominately coherent (classic) pulsators. We apply the model to the study of the Mira variable U~Per, and the semiregular variable L2~Pup and, following our classification, conclude that they are both classical pulsators. Our model provides a natural explanation for the change in behaviour of the pulsation amplitude-period relation noted in several earlier works. Moreover, our study of L2~Pup enables us to test the scaling relation between the mode line width and effective temperature, confirming that an exponential scaling reproduces well the data all the way from the main sequence to the asymptotic giant branch, down to temperatures about 1000~K below what has been tested in previous studies.}
\end{abstract}


\section{Introduction}
\label{introduction}
Stochastic noise from near-surface convection is generally accepted to be the main driver of the pulsations observed in solar-like pulsators, from the main-sequence to the red-giant evolution phase \citep[see, e.g][for a review]{houdek15}. However, the situation is not so clear for further evolved cool, luminous long-period pulsators located near the tip of the red-giant branch and in the asymptotic giant branch, that we shall collectively call the Long Period Variables (LPVs), comprising the semiregular and Mira variables. In fact, the driving mechanism behind the pulsations observed in LPVs, in particular in the semiregular variables, has been a matter of significant debate. On the theoretical side, stability studies of pulsations in LPVs face a significant challenge related to the need to correctly account for the coupling between the pulsations and convection. Several efforts to account for this coupling have been made over the years, generally pointing towards pulsations in LPVs being intrinsically unstable, with the turbulent pressure playing an important role in the driving of the pulsations \citep[e.g.][]{Gough67,munteanu05,xiong13,xiong18}. However, on the observational side, the similarity between the pulsation properties in semiregular variables and solar-like pulsators has led a number of authors to argue that pulsations in semiregular variables are stochastically excited \citep{JCD01,bedding05,Soszynski07,dziembowski10,mosser13,yu20}, as opposed to being coherently-driven and  intrinsically-unstable, as assumed for Mira stars. 

Independently of what is the main source of driving of pulsations in LPVs and of whether their modes are intrinsically stable or unstable, it is clear that convection interacts and influences the pulsations they exhibit. Indeed, even the Mira stars often show pulsation amplitude and phase variability which may be interpreted as  resulting from random fluctuations~\citep{eddington1929}, likely associated to the presence of convection~\citep{percy1999}. {This is in addition to secular variations observed in the pulsation periods of some Mira stars, whose origin may be related with the onset of thermal pulses \citep[e.g.][]{wood81,molnar19,templeton05}, as well as period variations observed on timescales of decades \citep[e.g.][]{Zijlstra02,templeton05}.} Recently, the impact of stochasticity on otherwise coherently-driven pulsators has been considered by \cite{avelino20} {(hereafter Paper~I)}, based on an internally driven damped harmonic oscillator model.

In the present paper we explore further the model introduced {in Paper~I}, by considering a range of possibilities for the impact of the stochastic perturbations on pulsations, from the limit when the pulsations are essentially stochastically driven to the limit when the coherent driving dominates. Since our model is phenomenological, we cannot draw conclusions on the exact physical sources of excitation and damping of the modes, as done in  stability analyses such as that discussed by \cite{xiong13}. However, our approach allows us to draw a bridge between theory and observations, by assuming that a pulsation mode is described by a model that incorporates simultaneously a coherent and a stochastic driving source, in addition to a damping term. The focus is on the model predictions for the power spectral density and how they compare with the known observational properties of pulsations in stars where these pulsations may be influenced by convection. 

The paper is organised as follows: in Sec.~\ref{sec:model} we briefly review the properties of the model introduced {in Paper~I} along with the main conclusions drawn in that work, which are of importance to the discussion presented here. Section~\ref{modifieda} discusses the properties of the power spectral density for a pulsation mode in the limit cases when stochasticity has no significant influence on the pulsation amplitude (classical limit) and when it dominates the driving (stochastic limit), as well as in the case when both drivings contribute significantly to the pulsation (intermediate case). In Sec.~\ref{test} we perform numerical simulations based on the proposed model to test the analytical predictions for the power spectral density introduced in Sec.~\ref{modifieda} and discuss further the impact of the main physical model parameters, namely, the mode lifetime and the amplitude of the stochastic noise, on the properties of the power spectral density. In Sec.~\ref{obs} we apply our model to the study of the Mira star U~Per and the semiregular star L2~Pup, and draw conclusions on the main driving source for the pulsations observed in these stars. In Sec.~\ref{discussion} we argue that our model provides a natural explanation to the change in behaviour observed in the pulsation amplitude-period relation of LPVs and show how the results of our analysis for L2~Pup allow us to test the mode line width-effective temperature scaling relation at low effective temperatures. Finally, in Sec.~\ref{conclusions} we summarise our main conclusions.

\section{Model}
\label{sec:model}
In this work we adopt the internally driven damped harmonic oscillator model {proposed \ACCnb}. {In this model the displacement, $x$, is described by the equation of motion 
\label{standard}
\label{}
\begin{equation}
{\ddot x\left(t\right)}+2 \eta {\dot x\left(t\right)} + \omega_0^2 x\left(t\right) = a_{f}\left(t\right) + \xi\left(t\right)\,, \label{fho}
\end{equation}
where a dot indicates a derivative with respect to time, $t$, $\eta$ is the damping constant, $\omega_0$ is the natural angular frequency of the oscillator, $a_f$ is the acceleration associated to the internal coherent driving mechanism, given by a sinusoidal function of frequency $\omega_f$ and a varying phase (see their section 3 for details), and $\xi$ is the function that parameterizes the stochastic noise.} The underlying assumption is that the stochastic perturbations induce a series of successive random velocity kicks separated by a time interval $\Delta t$. In that case, $\xi(t)$ may be written as
\begin{equation}
\xi(t) = 2 A_{N} \omega_0     \sqrt{\omega_0 \Delta t} \sum_{k=0}^{K} r(k) \, \delta(t-t_k)\,,
\label{xi1}
\end{equation}
where $A_N$ (with dimensions of a length) parameterizes the amplitude of the noise,  $r(k)$ are independent random variables with a normal distribution of mean zero and unit standard deviation ($r(k) \sim N(0,1)$), and $\delta$ is the Dirac delta function. Moreover, $t_k=k\Delta t$, and $t_{\rm obs}=K\Delta t$ is the total observing time. 
The normalisation adopted in Eq.~(\ref{xi1}) is such that the average velocity variation in a timescale equal to the oscillation period $P$ is proportional to $A_N \omega_0$.

The model described by Eq.~(\ref{fho}) will be used to study pulsations where stochastic and coherent driving are simultaneously present. There are two limits to this model, the first corresponding to setting $a_f=0$, in which pulsations are driven solely by stochastic perturbations, and the second corresponding to setting $\xi(t)=0$, in which the driving is fully coherent at all times. We shall refer to pulsations described by these limits as stochastic (or solar-like) and purely classical, respectively. In addition we shall introduce the classical limit (Sec.~\ref{class}), which is the limit when the impact of stochasticity on the pulsation amplitude is small, but the induced change in the pulsation phase can still be significant, if considered over a sufficiently long period of time.

In this work we are interested in bridging between the classical and stochastic limits, focusing, in particular, on how the pulsation power spectrum properties change as the impact of stochasticity increases. The classical limit has been studied {in detail \ACCnb}. In particular, the authors have shown that in this limit the phase variation displays a random walk behaviour, while the amplitude variation remains small. Working under the assumption that the observed (angular) frequency equals $\omega_f$ and is approximately equal to $\omega_0$ (requiring that $\eta^2\ll\omega_0^2/2$), the authors derived analytical expressions for the root mean square (rms) of the relative pulsation amplitude variation $\Delta A/A$ and of the phase variation $\Delta\varphi$ after a total observing time, $t_{\rm obs}$. These expressions, valid in the classical limit, are given by
\begin{equation}
\sigma_{\Delta A/A} =  \frac{A_N}{A} \sqrt{\frac{\omega_0}{\eta}}\,,
\label{siga_lim}
\end{equation}
and
\begin{equation}
\sigma_{\Delta \varphi} = \frac{A_N}{A} \sqrt{2\omega_0 t_{\rm obs}} \,,
\label{sigphi}
\end{equation}
respectively ({\it cf.} their equations 25 and 27), where $t_{\rm obs}$ is assumed to be much longer than $\eta^{-1}$. In the model $\Delta A$ and $\Delta\varphi$ are defined as the difference in amplitude and phase relative to the values in the case of no stochastic noise ($\xi=0$). In practice, when dealing with observations the exact underlying coherent amplitude of the pulsation is unknown. In that case, the time average of the amplitude, $\langle A\rangle$, can be used as an estimator of $A$ and $\Delta A$ computed with respect to that average. Moreover,  $\Delta\varphi$ can be computed with respect to any fixed phase value (e.g., the phase at the start of the observations). {Observationally, phase variations in variable-star research are usually expressed in terms of $O-C$ diagrams \citep[e.g.,][and references therein]{sterken05,catelan15}. In that context, $\Delta\varphi$ is to be interpreted as the stochastic signature seen in these diagrams. While the $O-C$ diagrams of some LPVs may also exhibit signatures of the decadal and secular pulsation period variations mentioned in Sec.~\ref{introduction}, these long-term variations will have a negligible impact on $\sigma_{\Delta \varphi}$, compared to the impact of the short-term stochastic variations}. In addition to characterising the phase and amplitude variations, the authors have shown that, unlike in solar-like pulsators, the power of the signal in the classical limit is independent of $\eta$.

Throughout the remainder of this paper, we will focus on the power spectrum distribution that results from the model described by Eq.~(\ref{fho}), considering not only the classical and stochastic limits, but also the intermediate regime in which the two driving terms have a comparable impact on the pulsations.



\section{Power spectrum distribution}
\label{modifieda}

\subsection{Classical limit}
\label{class}
When the coherent driving dominates, the amplitude variations remain small at all times, but the phase variation with respect to a fixed time (e.g., the time $t_0$ at the beginning of the observations) follows a random walk, its variance thus increasing with time. In this section we shall derive an analytical expression for the power spectrum of the signal, $x$, under the classical limit, which we define as being the limit when the amplitude variations may be neglected in the computation of the power (see \citet{Hajimiri99} for an analogous calculation in the context of circuits and communications). In that limit, and under the assumption that $\eta^2\ll\omega_0^2/2$ (so that the observed frequency is $\approx \omega_0$; {Paper~I}), the solution to Eq.~(\ref{fho}) can be written as
\begin{equation}
x=A \sin \left[\omega_0 t + \varphi\right]\,,
\end{equation}
where $A$ is essentially a constant (equal to the amplitude of the pulsation in the stochastic-free case) and $\varphi$ describes a random walk. 
 
 Let us start by defining the parameter 
\begin{equation}
\Gamma_{\rm c}\equiv \left(\frac{A_N}{A}\right)^2\frac{ \omega_0}{2\pi}=\frac{\sigma^2_{\Delta \varphi}}{4\pi \lvert\tau\rvert}\,,
\label{gammac}
\end{equation}
which is a constant in our problem. Here $\Delta \varphi[\tau] \equiv \varphi[t+\tau]- \varphi[t]$, $\tau$ being any given positive or negative time interval whose absolute value is much longer than the time interval between the random kicks $\Delta t$, and the equality follows directly from Eq.~(\ref{sigphi}), which is valid when the coherent driving dominates.

The variance of the phase change over an observation timescale $\tau$ is given by
\begin{equation}
\sigma^2_{\Delta \varphi}[\tau]=\langle ( \varphi [t+\tau]-\varphi [t])^2 \rangle\,,
\end{equation}
where the brackets represent an average over an infinite number of random walk realizations at fixed time $t$ (the time $t$ being arbitrary).

Next, consider the auto-correlation function of the signal given by
\begin{eqnarray}
&& \hspace{-1cm}\langle x [t] x [t+\tau] \rangle\nonumber = \\
&=& A^2 \langle \sin \left[\omega_0 t + \varphi[t] \right] \sin \left[\omega_0 (t+\tau) + \varphi[t+\tau] \right] \rangle\nonumber\\
&=& \frac{A^2}{2} \langle \cos \left[\omega_0 \tau + \Delta \varphi \right]\rangle \nonumber\\
&-& \frac{A^2}{2} \langle \cos \left[\omega_0 (2 t+\tau) + 2\varphi[t]+ \Delta \varphi\right] \rangle\,. \label{auto1}
\end{eqnarray}
 If the time $t$ is sufficiently long, the last term in Eq. \eqref{auto1} vanishes. Indeed, due to its random walk nature, the phase $\varphi$ 
in the last term of Eq.~\eqref{auto1} will follow a normal distribution. When the time $t$ at which the random walk realisations are considered is sufficiently long to ensure that the rms of $\varphi$ is much greater than 2$\pi$, $\varphi\, {\rm mod}\, 2\pi$ will approach a uniform distribution in the interval [0,2$\pi$] and the average of the last term in Eq.~\eqref{auto1} over many realisations will tend to zero.   Hence, the auto-correlation function becomes simply
\begin{eqnarray}
\langle x [t] x [t+\tau] \rangle &=&  \frac{A^2}{2} \langle \cos \left[\omega_0 \tau + \Delta \varphi \right]\rangle \nonumber\\
&=& \frac{A^2}{2}  \cos \left[\omega_0 \tau\right] \langle \cos \Delta \varphi \rangle \\
&-& \frac{A^2}{2}  \sin \left[\omega_0 \tau\right] \langle \sin \Delta \varphi \rangle \,, \label{auto2}
\end{eqnarray}

The probability density distribution of the phase variation $\Delta \varphi[\tau]$ on a timescale $\tau$, is given by
\begin{equation}
p(\Delta \varphi)=\frac{1}{\sqrt{2\pi} \sigma_{\Delta \varphi}} \exp\left(-\frac{\Delta \varphi^2}{2 \sigma^2_{\Delta \varphi}}\right)\,.
\end{equation}
Therefore
\begin{eqnarray}
\label{cosav}
\langle \cos \Delta \varphi \rangle &=& \int_{-\infty}^{\infty} p[\Delta \varphi] \cos[\Delta \varphi] d \Delta \varphi\label{avcos}\\
&=&\exp\left[ -\frac{\sigma^2_{\Delta \varphi}}{2}\right]=\exp\left[ -2 \pi \Gamma_c\lvert\tau\rvert\right]\nonumber\\
\langle \sin \Delta \varphi \rangle &=& \int_{-\infty}^{\infty} p[\Delta \varphi] \sin[\Delta \varphi] d \Delta \varphi=0 \label{sinav}\,,
\end{eqnarray}
where the last equality in Eq.~\eqref{avcos} follows from Eq.~\eqref{gammac}.

Substituting Eqs~(\ref{cosav}) and (\ref{sinav}) into Eq. (\ref{auto2}), the auto-correlation function may be finally written as
\begin{equation}
\zeta(\tau)=\langle x [t] x [t+\tau] \rangle=\frac{A^2}{2} \cos[\omega_0 \tau] \exp\left(- 2 \pi \Gamma_{\rm c} \lvert\tau\rvert \right)\,.
\end{equation}
{The limit power spectral density PSD ({\itie} the power spectral density expected after an infinite number of realisations)} may then be obtained by performing the Fourier transform\footnote{here and throughout this work we have adopted the following definition for the Fourier transform of a function $g(t)$: $G(\omega)=\int_{-\infty}^{+\infty} e^{-i\omega t} g(t) dt$} of the auto-correlation function of the signal. Defining $f_0=\omega_0/2\pi$ and $f=\omega/2\pi$, we have
\begin{eqnarray}
&& \hspace{-1cm}\rmpsd(f)=\int_{-\infty}^{+\infty} e^{-i\omega \tau} \zeta(\tau) d \tau = \nonumber\\
 && \hspace{-1cm} =\frac{A^2}{4 \pi \Gamma_c} \left(\frac{1}{1+\left(f-f_0\right)^2/\Gamma_c^2}+\frac{1}{1+\left(f+f_0\right)^2/\Gamma_c^2}\right) \,,
\end{eqnarray}
In general $\Gamma_c \ll f_0$, thus implying that the limit power spectral density around the frequency $f_0$ (for $|f-f_0| \ll f_0$) is well approximated by a Lorentzian function. \\

{This power is to be compared with that computed from the light curves of pulsating stars. Hence we shall consider a single-sided power spectrum and, accordingly, multiply the power in each positive frequency by two, to ensure it is calibrated according to Parseval’s Theorem.  Doing so, the power spectral density around the frequency $f_0$ in the classical limit becomes }
\begin{equation}
\rmpsd(f) = \frac{A^2}{2 \pi \Gamma_{\rm c}} \frac{1}{1+\left(f-f_0\right)^2/\Gamma_c^2} \,,
\label{lor_clas}
\end{equation}
representing a Lorentzian function with a half width at half maximum power equal to $\Gamma_{\rm c}$. Thus, so long as a classical pulsator in the presence of stochastic noise is observed for a long enough period of time, we may expect the power spectral density of a single pulsation mode to be a resolved Lorentzian function whose width depends on the amplitude of the stochastic noise {(tending to zero in the limit when ~$\xi\to 0$)} but not on the damping constant. 


\subsection{Stochastic limit}
\label{sec:stochastics}

The opposite limit is the one in which the driving is dominated by the stochastic source and the amplitude of the coherent oscillations $A \to 0$. In this limit the equation of motion describing the evolution of $x$ with time approaches that of a damped harmonic oscillator with noise as given by 
\begin{equation}
{\ddot x}+2 \eta {\dot x} + \omega_0^2 x = \xi\,. \label{fhos}
\end{equation}
The Fourier transform of Eq. (\ref{fhos}) is given by
\begin{equation}
-\omega^2 {\tilde x} - 2 i \eta \omega {\tilde x} + \omega_0^2 {\tilde x} = {\tilde \xi}  \,, \label{fho1}
\end{equation}
where a tilde indicates the Fourier transform of the corresponding function.

The function $\xi$ is defined by Eq.~\eqref{xi1}. Its Fourier transform is then
\begin{eqnarray}
{\tilde \xi}(\omega) &=&  2 A_{N} \omega_0     \sqrt{\omega_0 \Delta t}  \sum_{k=0}^K r(k) \, \int_{-\infty}^{\infty} \delta(t-t_k) e^{-i\omega t} dt \nonumber\\
&=& 2 A_{N} \omega_0     \sqrt{\omega_0 \Delta t} \sum_{k=0}^{K} r(k) e^{-i \omega t_k} \,.
\end{eqnarray}
Hence, the limit power spectral density of $\xi$ is given approximately by
\begin{equation}
\rmpsd_\xi(\omega)= \frac{\langle|\tilde \xi|^2\rangle}{T} = 4 A_{N}^2 \omega_0^3 \,,
\label{psdxi}
\end{equation}
where we have taken into account that
\begin{equation}
\sum_{k=0}^{K} r(k)^2 \sim K = \frac{T}{\Delta t} \,,
\end{equation}
for $K \gg 1$.

The limit power spectral density of the signal, $x$, expressed in terms of the angular frequency $\omega$, can then be derived from Eqs~\eqref{fho1} and \eqref{psdxi}, being given by
\begin{eqnarray}
\label{SD}
\rmpsd(\omega)&=&\frac{\langle|\tilde x|^2\rangle}{T}=\frac{1}{T} \frac{\langle|\tilde \xi|^2\rangle}{(\omega_0^2-\omega^2)^2 + 4\eta^2  \omega^2} \nonumber \\
&\approx&  \frac{A_N^2\omega_0}{(\omega_0-\omega)^2 +  \eta^2}
\end{eqnarray}
where we have taken into account that $(\omega_0^2-\omega^2)^2 \sim 4 \omega_0^2(\omega-\omega_0)^2$ for $\omega \sim \omega_0$, up to first order in $\omega-\omega_0$.  

Taking into account that $\omega=2 \pi f$, and binning the power on negative frequencies onto their positive counterparts (as done in Sec.~\ref{class}), one finally obtains the limit power spectral density expressed in terms of the cyclic frequency,
\begin{equation}
\label{Px}
\rmpsd(f) =  \frac{A^2 \Gamma_{\rm c}}{\pi \Gamma_{\rm s}^2} \frac{1}{1+\left(f-f_0\right)^2/{\Gamma_{\rm s}}^2}\,,
\end{equation}
where 
\begin{equation}
\Gamma_s=\eta/(2\pi)\,, \qquad \Gamma_c=\left(\frac{A_N}{A}\right)^2\frac{ \omega_0}{2\pi}\,.
\label{gammas}
\end{equation}
In writing Eq. (\ref{Px}) we have assumed that $A$ is small but non-zero. If $A$ is exactly zero then the prefactor in Eq. (\ref{Px}) should be written as $A_N^2 \omega_0/(2\pi^2 \Gamma_s^2)$. In the stochastic limit we therefore recover the well known result that the limit power spectral density of a single pulsation mode is described by a Lorentzian function whose width depends only on the damping constant and not on the amplitude of the stochastic noise.


\subsection{Intermediate regime}
\label{interm}
According to the model described by Eq.~\eqref{fho}, the limit power spectral density for a single pulsation mode is well described by a Lorentzian function in both the classical (Sec.~\ref{class}) and stochastic (Sec.~\ref{sec:stochastics}) limits. In the intermediate regime, where both the coherent and stochastic driving play a role, the assumptions underlying the analytical derivations break down and the problem must be addressed numerically. Nevertheless, it would be of interest to have an approximate analytical solution that could be used to estimate the main underlying physical parameters of the model,  $\eta$ and $A_N/A$, in the intermediate regime, in addition to the mode frequency. 

To achieve that, let us start by defining
\begin{equation}
\Gamma=\left(\Gamma_{\rm c}^{-1}+\Gamma_{\rm s}^{-1}\right)^{-1} \,,
\label{total_gamma}
\end{equation}
so that $\Gamma=\Gamma_{\rm c}$ if $\Gamma_{\rm c} \ll \Gamma_{\rm s}$ (classical limit) and $\Gamma=\Gamma_{\rm s}$ if $\Gamma_{\rm s} \ll \Gamma_{\rm c}$ (stochastic limit) (note that $\Gamma \le \Gamma_c, \Gamma_s$). With this definition in hand, it is possible to write a single expression for the power spectral density that is valid both in the classical and stochastic limits, namely,
\begin{equation}
\rmpsd(f) =  \frac{A^2}{2\pi \Gamma}\left(1+2 \frac{\Gamma_{\rm c}}{\Gamma_{\rm s}} \right) \frac{1}{1+\left(f-f_0\right)^2/\Gamma^2}\,.
\label{intermediate}
\end{equation}
As we will show in Sec.~\ref{test}, this expression constitutes also a good approximation to the limit power spectral density of a pulsation mode in the intermediate regime where $\Gamma_{\rm s}/\Gamma_{\rm c} \sim 1$. Consequently, Eq.~\eqref{intermediate} provides a {unifying} description of the limit power spectra density of individual modes in stellar pulsators where stochastic perturbations play a role.

\section{Test on numerical simulations}
\label{test}
Equation~\eqref{intermediate} reproduces exactly the limit power spectral density predicted in the classical (Eq.\eqref{lor_clas}) and stochastic (Eq.\eqref{Px}) limits. To verify that it also provides a good representation of the limit power spectral density for cases when both the classical and stochastic drivings impact the oscillations, we carried out a number of numerical simulations based on Eq.~\eqref{fho}. Specifically, we numerically solved the equation of motion using a fourth order Runge-Kutta algorithm, setting the initial conditions to be those appropriate for the purely classical pulsations ({\it i.e.,} the solution for $\xi=0$). In the simulations, time is measured in units of the pulsation period $P$ and length in units of the amplitude $A$. The time interval between successive kicks was assumed to be $\Delta t = 0.025P$. The fact that it is much shorter than the pulsation period ensures that both the assumed regularity of the velocity kicks and the specific value of the parameter $\Delta t$ have a negligible impact on our results. 

Each realisation of the simulations spans a total time equal to $5\times 10^4$ pulsation cycles. We consider two series of simulations, one with $\eta=0.05P^{-1}$ and the other with $\eta=0.5P^{-1}$. Since $\omega_0\approx2\pi P^{-1}$, the condition that $\eta^2 \ll \omega^2/2$ (Sec.~\ref{class}) is satisfied in all cases. For each series, we consider different values of $A_N/A$ such that the ratio $\Gamma_c/\Gamma_s$ changes from a value of 10000 to a value of 0.01. The aim is to illustrate the changing in regime, from a stochastic dominated case (where $\Gamma_c/\Gamma_s > 1$) to a classical dominated case (where $\Gamma_c/\Gamma_s< 1$). 

\begin{figure*} 
\noindent\makebox[0.49\textwidth][r]{%
	\begin{minipage}{0.45\linewidth}  
		\rotatebox{0}{\includegraphics[width=1.0\linewidth]{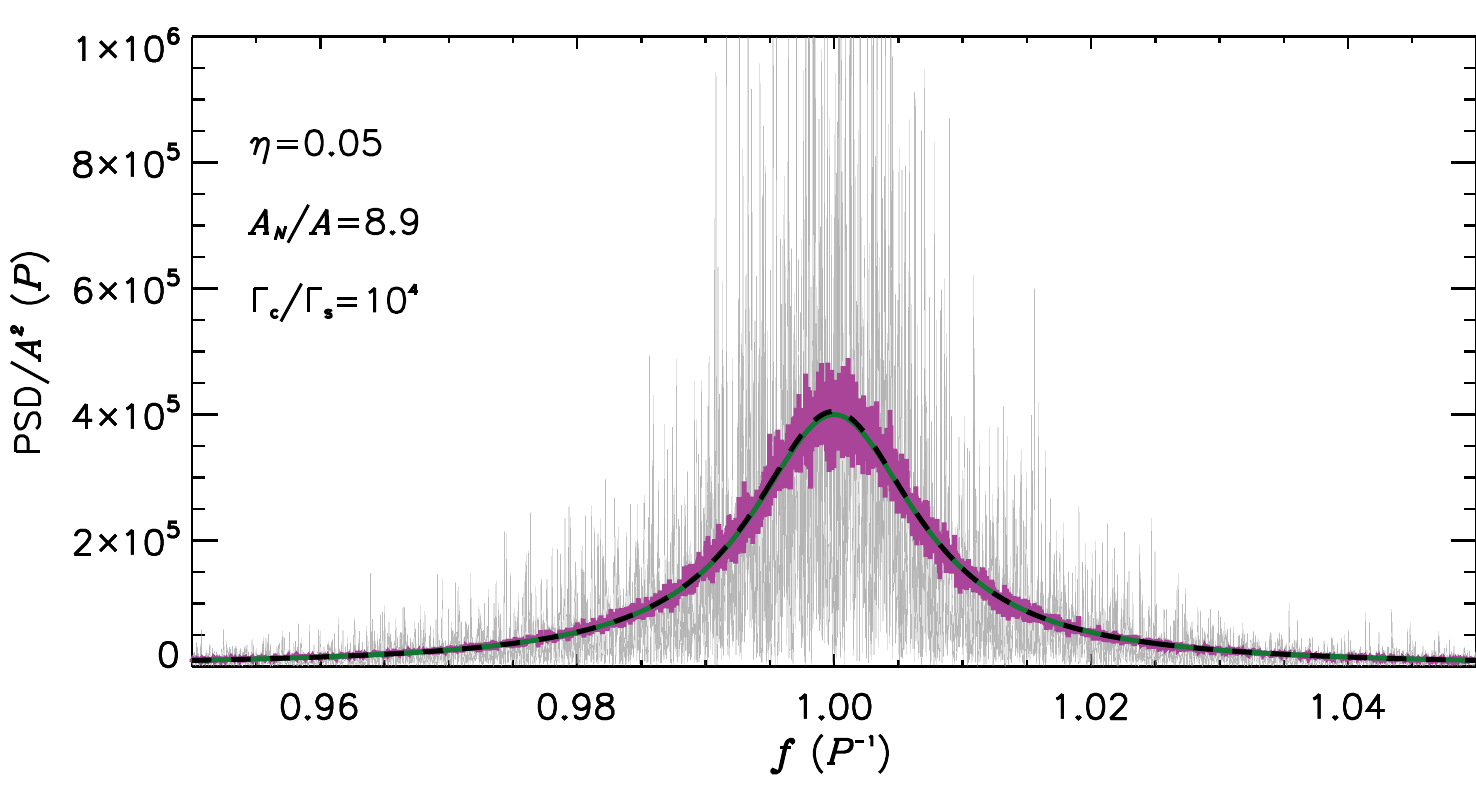}} 

		\rotatebox{0}{\includegraphics[width=1.0\linewidth]{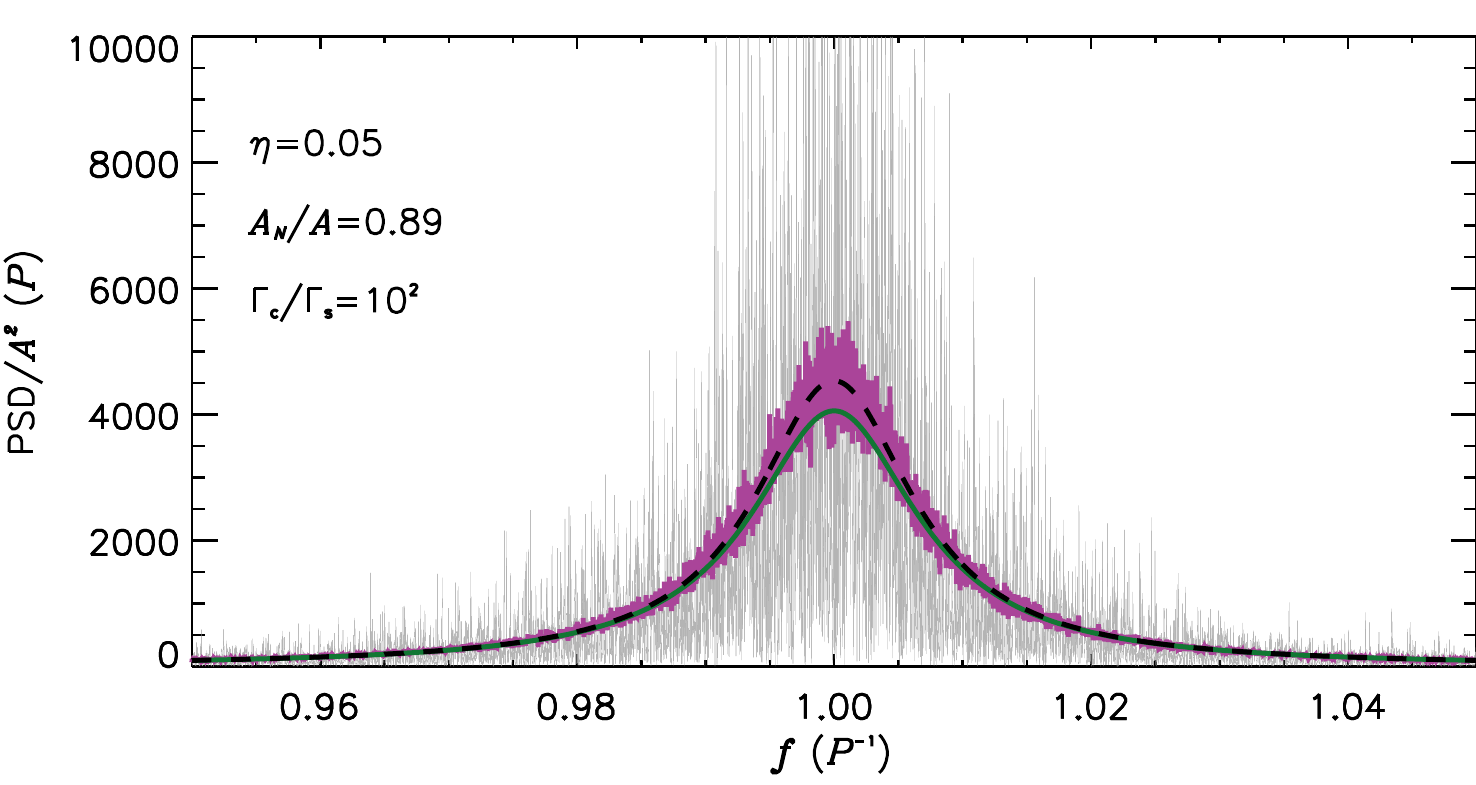}}
		
		\rotatebox{0}{{\includegraphics[width=1.0\linewidth]{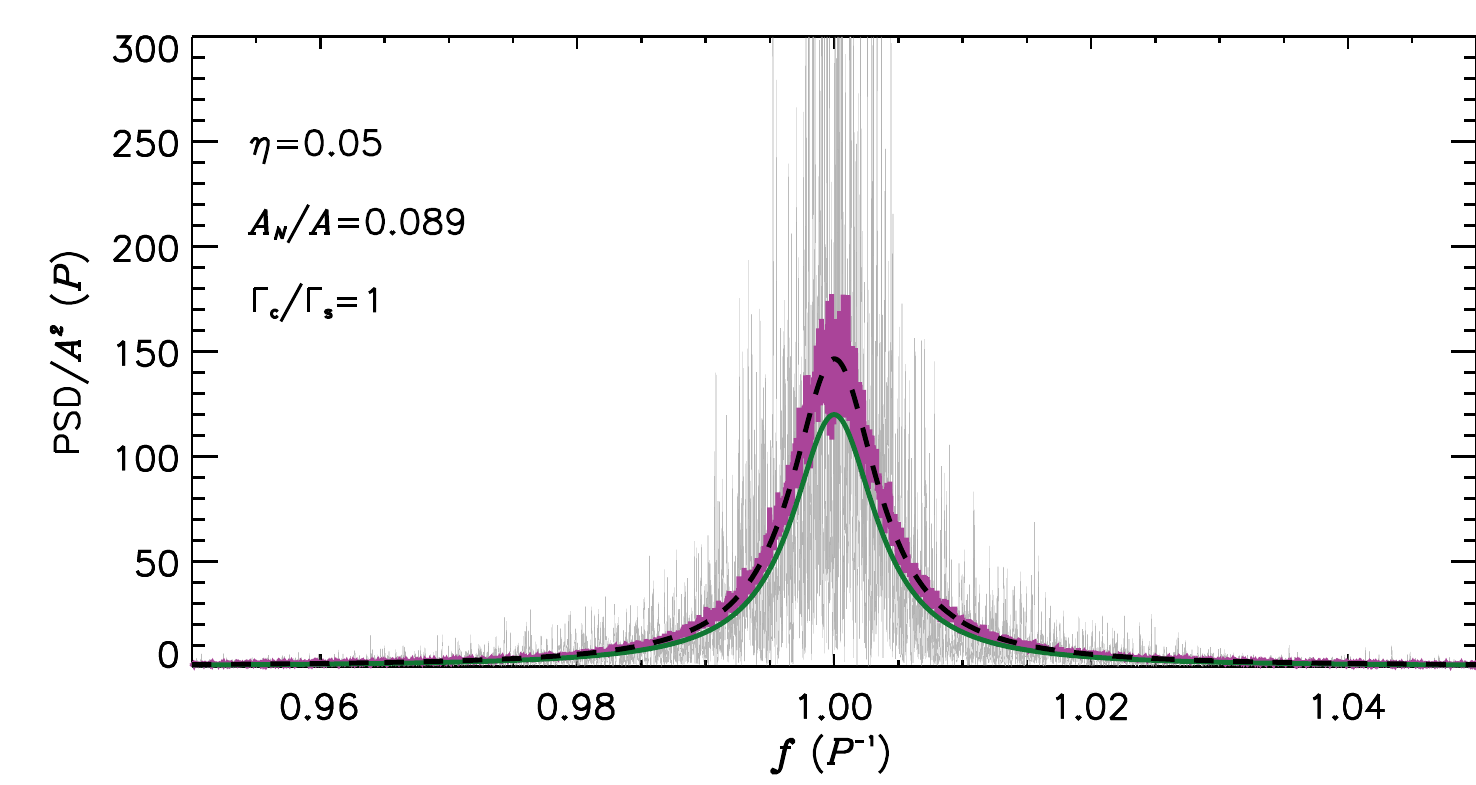}}}

		\rotatebox{0}{\includegraphics[width=1.0\linewidth]{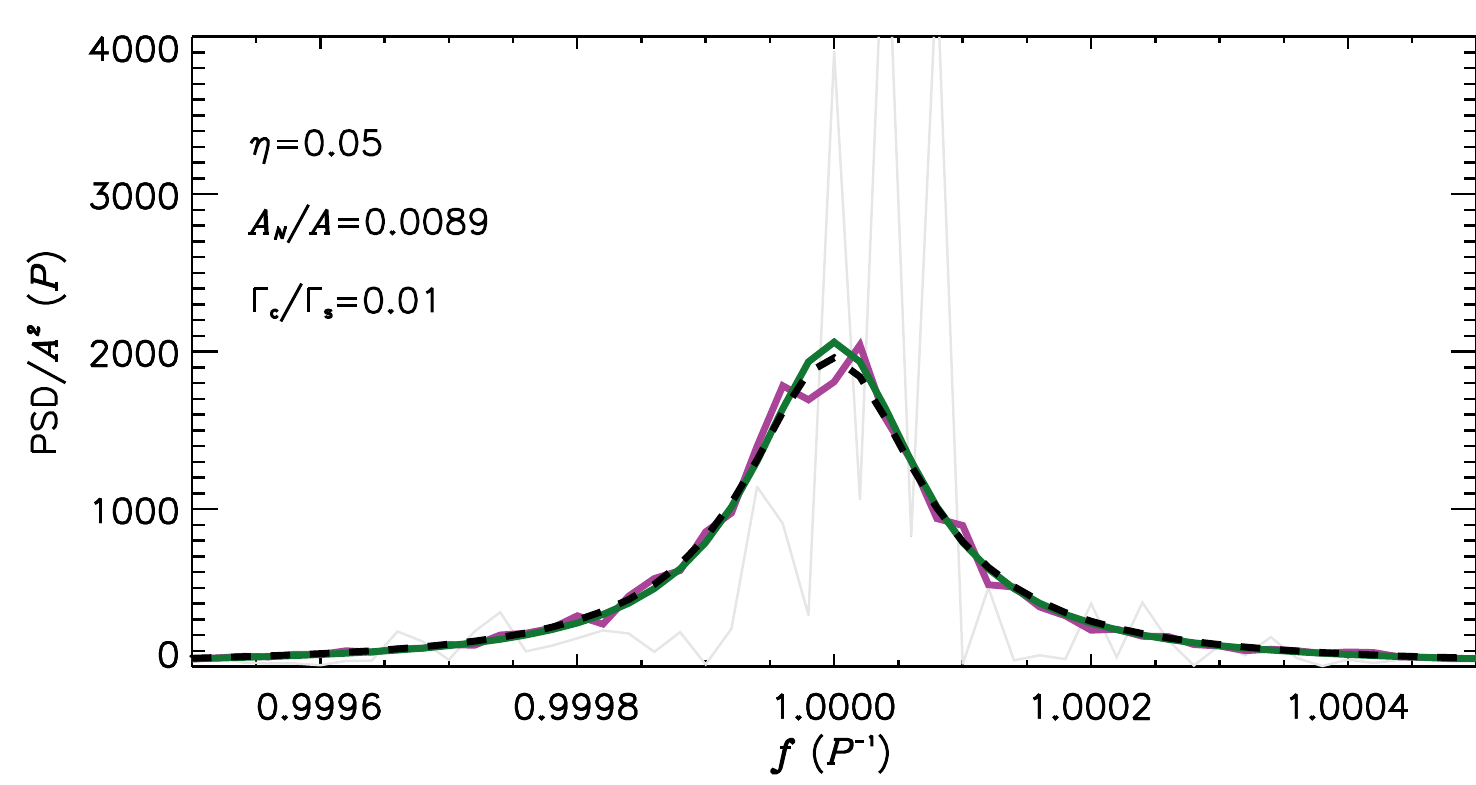}}
	\end{minipage}
	}
\noindent\makebox[0.49\textwidth][r]{%
		\begin{minipage}{0.45\linewidth}  
		\rotatebox{0}{\includegraphics[width=1.0\linewidth]{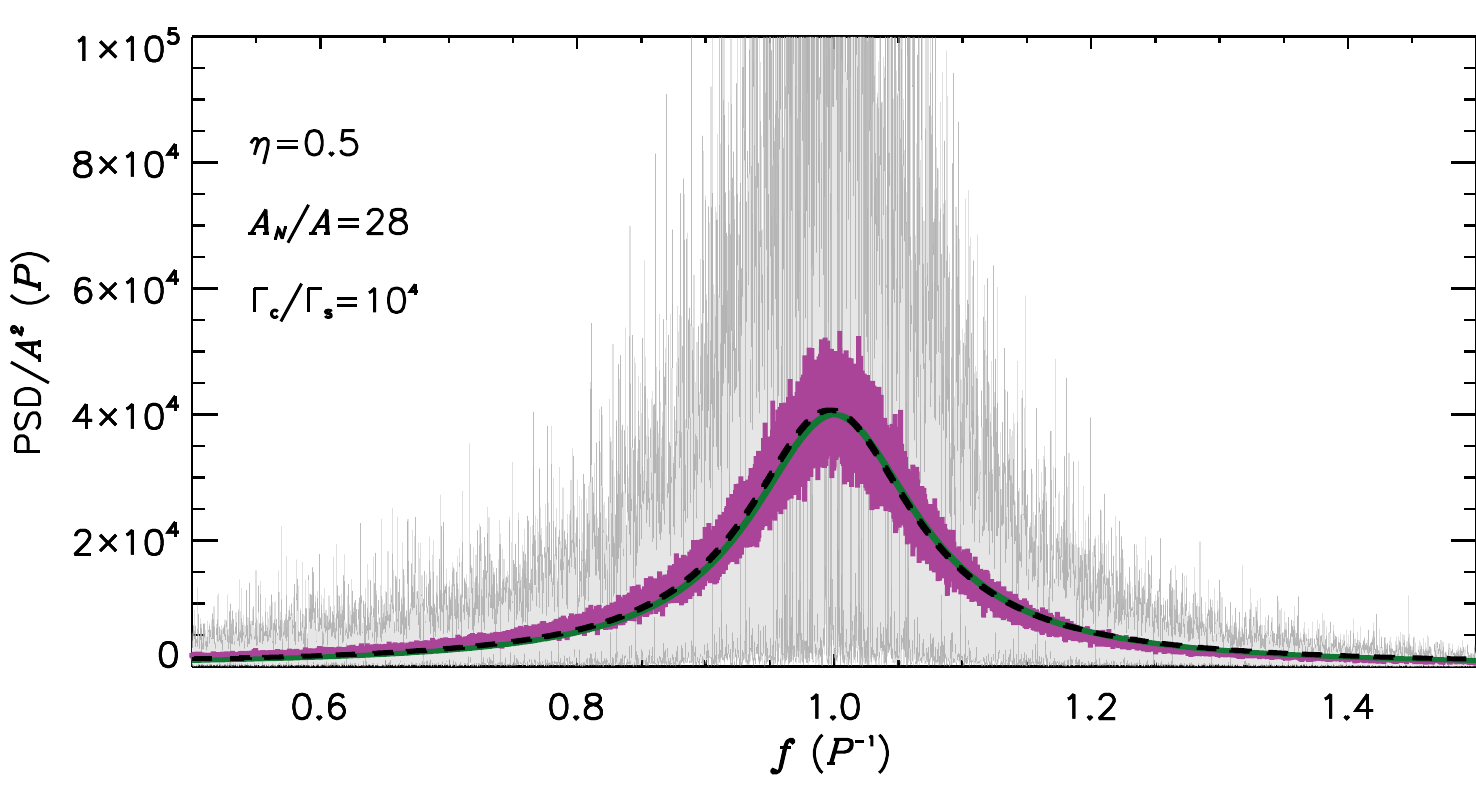}}

		\rotatebox{0}{\includegraphics[width=1.0\linewidth]{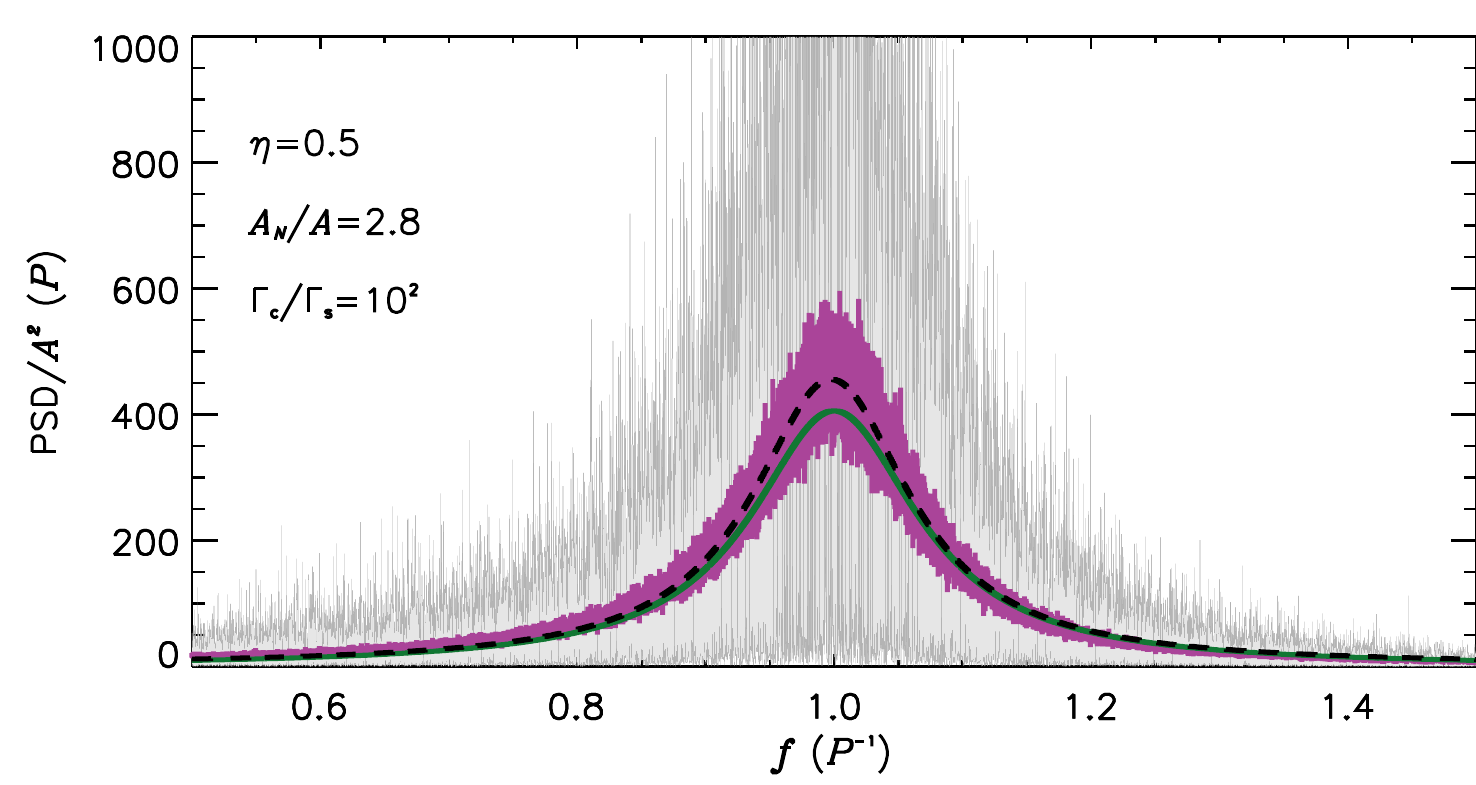}}

		\rotatebox{0}{{\includegraphics[width=1.0\linewidth]{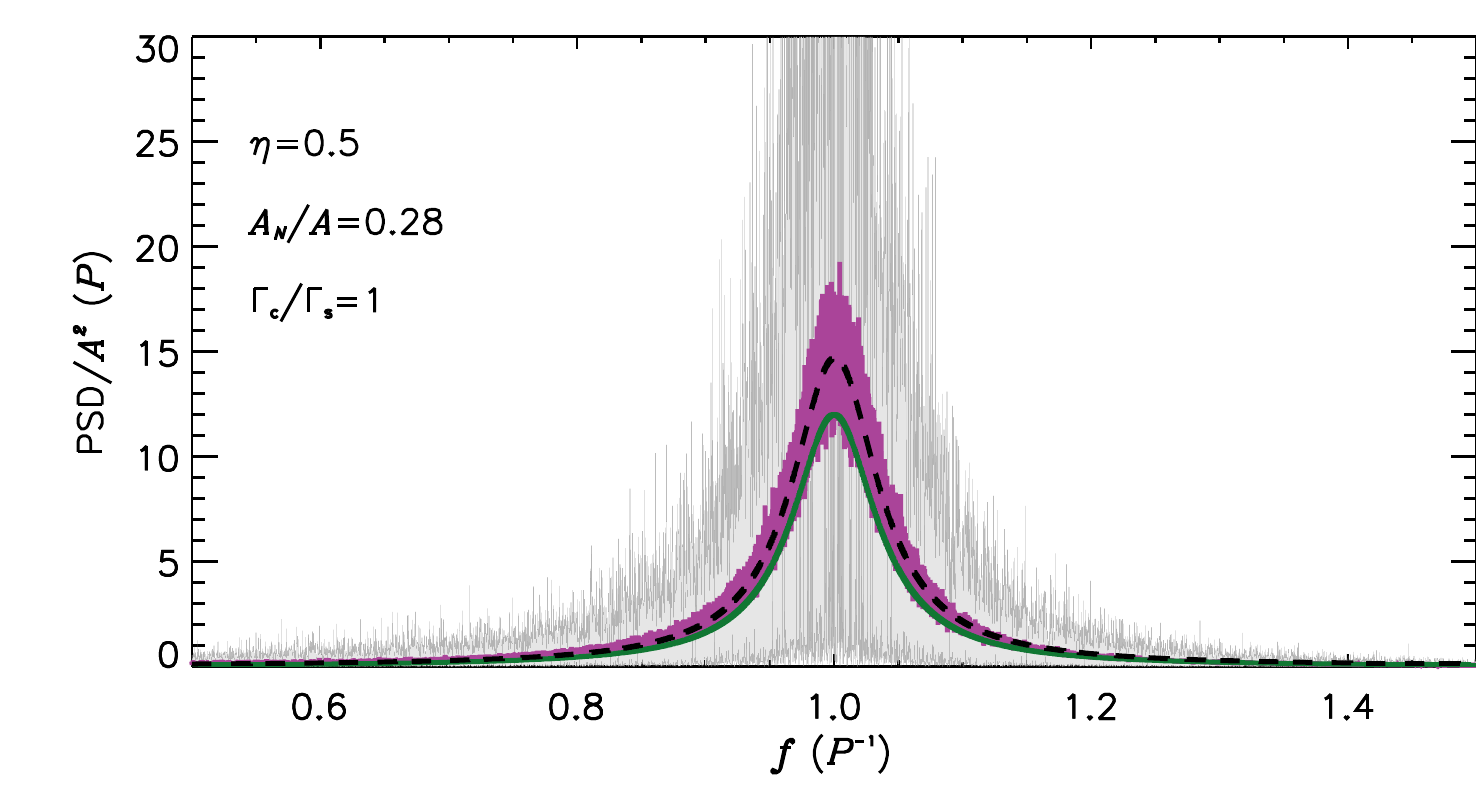}}}

		\rotatebox{0}{\includegraphics[width=1.0\linewidth]{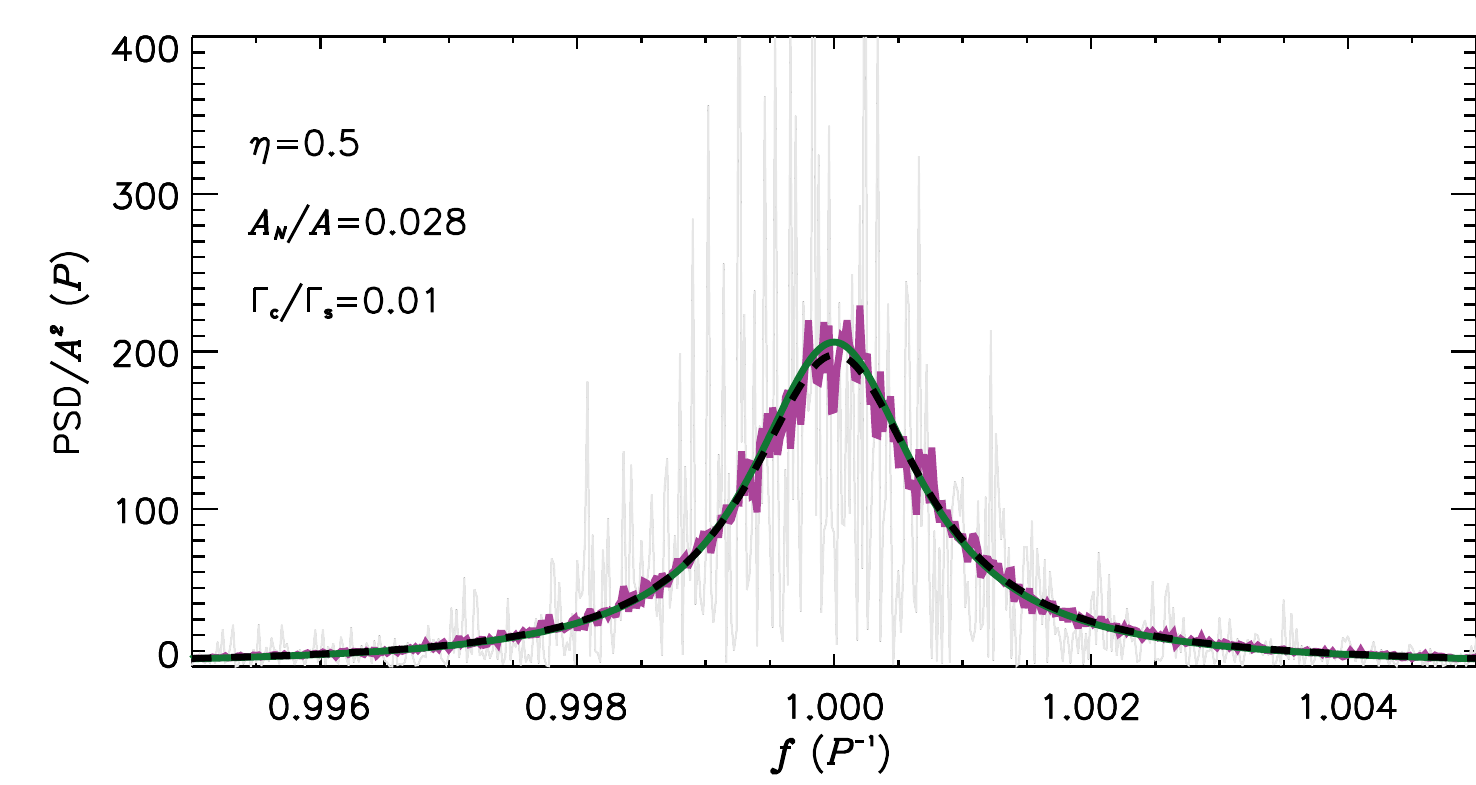}}
	\end{minipage}
	}

	\caption{Amplitude-normalised power spectral density for numerical simulations based on the model described by Eq.~(\ref{fho}). Panels in the left and right columns display results for $\eta=0.05P^{-1}$ and  $\eta=0.5P^{-1}$, respectively.  The value of $A_N/A$ adopted in each simulation is shown in the inset of the corresponding panel, along with the value of $\eta$ in units of $P^{-1}$. The ratio $\Gamma_c/\Gamma_s$ derived from these parameters through Eq.~(\ref{gammas}) is also shown. {\bf Light grey line:} one realisation of the simulations; {\bf Pink line:} the average of 100 realisations; {\bf Dashed-black line:} the Lorentzian function derived from fitting the average shown in pink; {\bf Green line:} the Lorentzian function given by Eq.~(\ref{intermediate}) computed with the parameters of the corresponding simulation. The transition from a stochastically-dominated model (top panels) gradually to a classically-dominated model (bottom panels), is evident from the change in the width of the Lorentzian function at fixed $\eta$.
	}
	\label{fig:simulations}
\end{figure*}

The results of our simulations are displayed in Fig.~\ref{fig:simulations}, where the series with $\eta=0.05P^{-1}$ and $\eta=0.5P^{-1}$ are shown on the left and right columns, respectively. For each simulation we show one single realisation (light grey line) and the average of one hundred realisations (pink line). In addition, we show the predicted limit power spectral density, given by Eq.~\eqref{intermediate} (green line), and the Lorentzian function derived from a fit to the average power spectral density (black-dashed line). 
{All results are shown in the form of a power spectral density divided by the square of the underlying coherent pulsation amplitude ({\it i.e.} PSD/$A^2$). Besides the dependence on $f_0$, the quantity PSD/$A^2$ (hereafter the amplitude-normalised power spectral density) depends only on the parameters $\Gamma_{\rm c}$ and $\Gamma_{\rm s}$, whose impact on the simulations we want to explore.}

As expected, when the driving is strongly dominated by either the stochastic or the coherent source (upper and lower panels in Fig.~\ref{fig:simulations}, respectively),  Eq.~\eqref{intermediate} provides an excellent description of the average power spectral density obtained from our simulations. In the intermediate cases (four middle panels in Fig.~\ref{fig:simulations}), Eq.~\eqref{intermediate} predicts an amplitude that is somewhat smaller than that found in our simulations, but a half width at half maximum power always very close to that of the simulations. Comparing the parameters characterising the Lorentzian function derived from the fit to the average power spectral density and the theoretical Lorentzian given by Eq.~\eqref{intermediate}, we find a maximum relative difference in the amplitude of 22$\%$ (for the cases with $\Gamma_c/\Gamma_s=1$). Moreover, for a fixed value of the ratio $\Gamma_c/\Gamma_{\rm s}$, we find that this difference is the same (to better than one per cent) for the two values of $\eta$ considered. For the half width at half maximum power, the agreement is always better than 5$\%$. 

The dependence of the width of the Lorentzian describing the power spectral density on the model parameters is also clear from Fig.~\ref{fig:simulations}. Comparison of the two top panels confirms the well known result that in the stochastically dominated case the width is proportional to $\eta$ (note the factor of 10 difference in the range of the x-axis between the two panels). However, when comparing the three top panels on either column, we can see that the width of the Lorentzian decreases (at fixed $\eta$) as $\Gamma_c/\Gamma_s$ decreases (i.e, as the impact of the coherent driving becomes non-negligible). Finally, when the coherent driving dominates (bottom panels), the width becomes independent of $\eta$ (see figure 5 of {Paper~I}, and Sec.~\ref{class} above). Here, the factor of 10 change in width seen between the left and right bottom panels (note the scale on the x-axis) is due to the change of $A_N/A$ and the fact that at fixed pulsation frequency $\Gamma_c$ is proportional to $(A_N/A)^2$. Finally, we note that the factor of 10 change in the amplitude between the left and right panels is a natural consequence of our choice of changing $\eta$ by a factor of 10 at fixed $\Gamma_c/\Gamma_s$. Increasing $\eta$ by one order of magnitude increases $\Gamma_s$ by the same factor, hence, at  fixed $\Gamma_c/\Gamma_s$, also $\Gamma_c$. This, in turn, increases $\Gamma$ by an order of magnitude (\itcf Eq.~\eqref{total_gamma}) reducing the amplitude by a factor of 10 (\itcf Eq.~\eqref{intermediate}).


\section{Observational tests}
\label{obs}
In this section we shall verify that the analytical expression for the power spectral density  of the signal derived in Sec.~\ref{modifieda} (Eq.~\eqref{intermediate}) represents well the observations of long-period variables. To that end, we will consider two stars: the Mira variable U~Per and the semiregular variable L2~Pup. As we shall see, these stars provide good examples of the classical limit (Sec.~\ref{class}) and of the intermediate regime (Sec.~\ref{interm}), respectively. The stochastic limit (Sec.~\ref{sec:stochastics}) has been extensively discussed in the literature, in connection to the study of solar-like pulsators \cite[e.g.][and references therein]{houdek19,belkacem19}, thus it shall not be of further concern here.

\subsection{Data}
\begin{figure} 
	\begin{minipage}{1.\linewidth}  
		\rotatebox{0}{\includegraphics[width=1.0\linewidth]{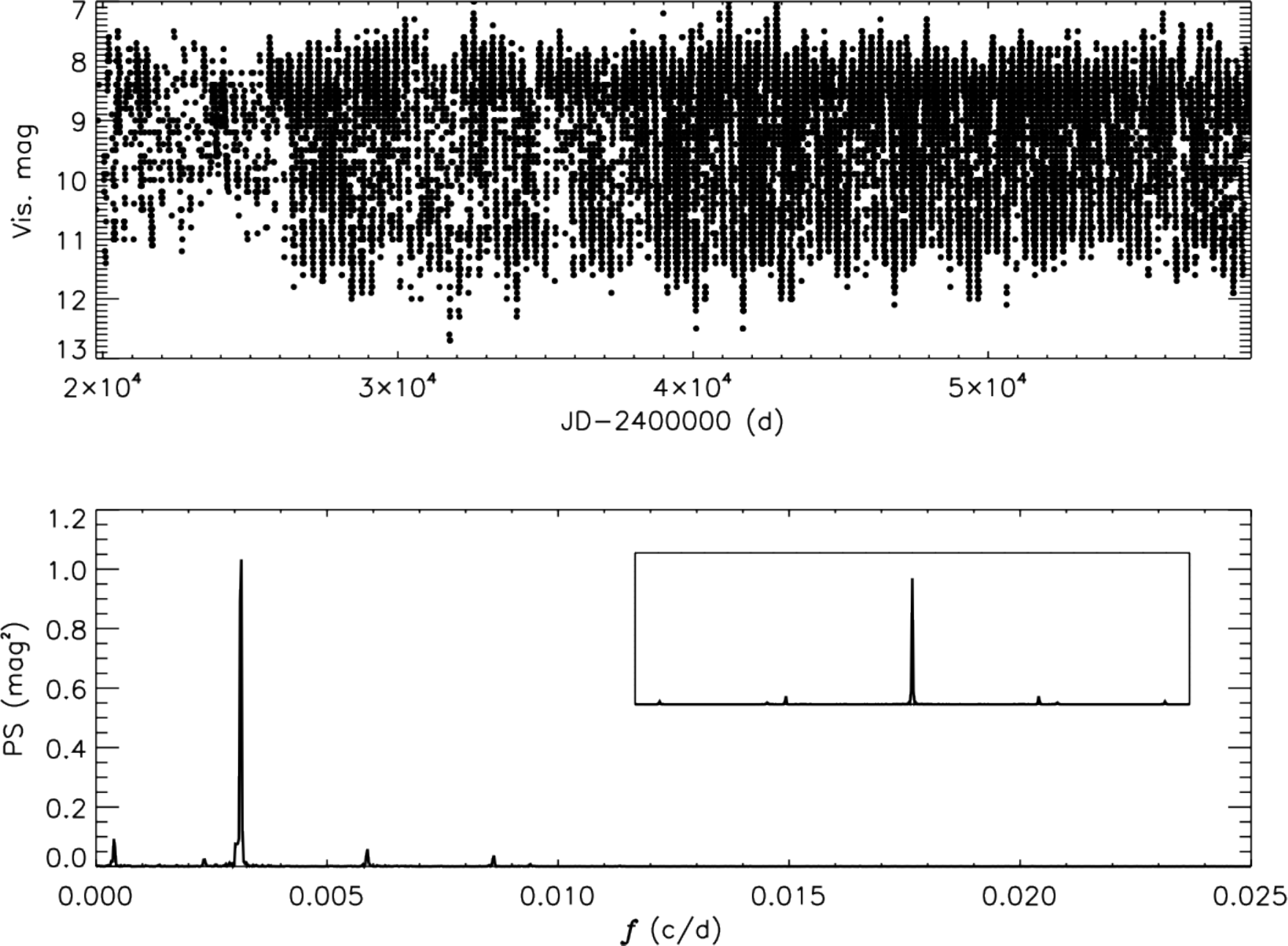}}
	\end{minipage}
	\caption{Top: {U Per light curve extracted from AAVSO. The data points are provided rounded to the decimal point, which explains the discrete appearance of the light curve.} Data runs from JD 2419771.74 to JD 2458909.522 in a total of 26525 data points collected over 107 years. Bottom: Power spectrum of the light curve showing the star's pulsation frequency near 0.0031~c/d along with side lobes at $\pm$~1 and 2~cycles per year ({\it cf.} window function shown in the inset).
	}
	\label{fig:uper}
\end{figure}

\begin{figure} 
	\begin{minipage}{1.\linewidth}  
		\rotatebox{0}{\includegraphics[width=1.0\linewidth]{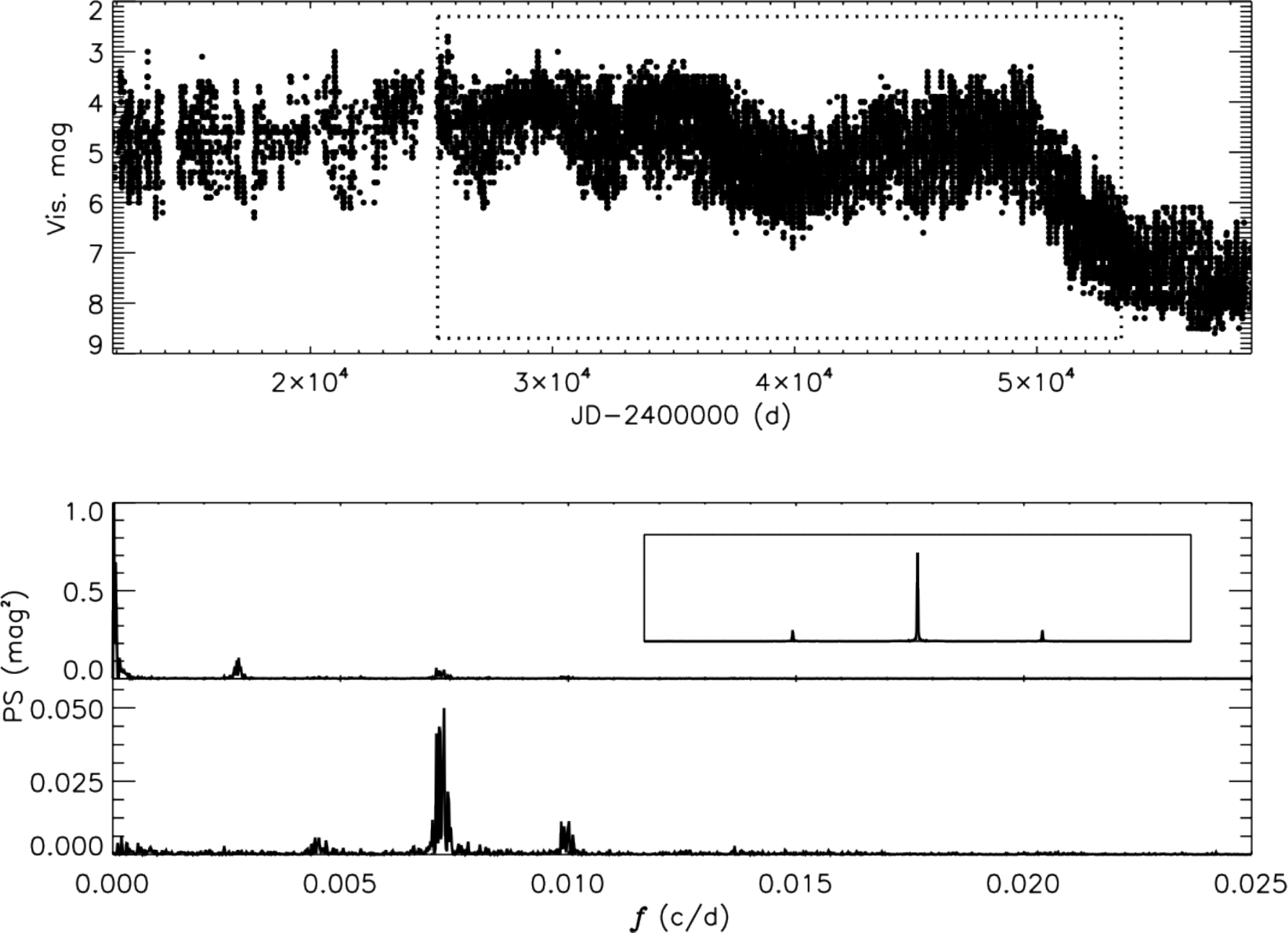}}
	\end{minipage}
	\caption{Top: L2 Pup light curve. Data runs from JD 2411839.4 to JD 2458870.074 in a total of 21448 data points collected over nearly 129 years. The dotted line shows the observation period analysed by \citet{bedding05}, running from JD 2425249 to JD 2453487, and considered for comparison in Sec.~\ref{l2pup}. Middle: Power spectrum of the light curve dominated by low frequency power associated to the long-period modulation of the light curve and the correspondent 1 cycle per year side lobe ({\it cf.} window function shown in the inset). Bottom: Power spectrum after pre-whitening by the low frequencies, showing the star's pulsation frequency near 0.007~c/d and the $\pm$~1~cycle per year side lobes.}
	\label{fig:L2pup}
\end{figure}

We downloaded visual magnitudes for U~Per and L2~Pup from the American Association of Variable Star Observers (AAVSO) data base\footnote{Kafka, S., 2020, Observations from the AAVSO International Database, https://www.aavso.org}. The corresponding light curves are shown on the top panels of Figs~\ref{fig:uper} and \ref{fig:L2pup}, respectively.  To identify the pulsation periodicity, we performed a Fourier analyses of the data using Period04 \citep{lenz05}. The power spectra of the light curves  are shown in the bottom panel of Fig.~\ref{fig:uper} and middle panel of Fig.~\ref{fig:L2pup}, respectively. The pulsation frequency of U~Per is clearly visible in the power spectrum, along with side lobes at at $\pm$~1 and 2 cycles per year (0.00273973 c/d). However, in the case of L2~Pup the power spectrum is dominated by low frequency signals that result from the long period amplitude modulation of the light curve. We have used Period04 to iteratively pre-whiten the data to remove the low frequency signal in the frequency interval 0-10$^{-4}$~c/d. The power spectrum of the residuals is shown in the bottom panel of Fig.~\ref{fig:L2pup}, where the power is now dominated by the pulsation frequency and its $\pm$~1 cycle per year side lobes.




\subsection{The Mira variable U Per}
\label{uper}
The data considered for U~Per extends for over 107 years (from JD 2419771.74 to JD 2458909.522), covering nearly 123 pulsation cycles. A clear pulsation amplitude variability is observed in the star's light curve shown on the top panel of Fig.~\ref{fig:uper}. Here we propose that this variability is due to stochastic perturbations that can be well described by the phenomenological model introduced {in Paper~I} and further discussed in this paper. 

\begin{figure} 
	\begin{minipage}{1.\linewidth}  
		\rotatebox{0}{\includegraphics[width=1.0\linewidth]{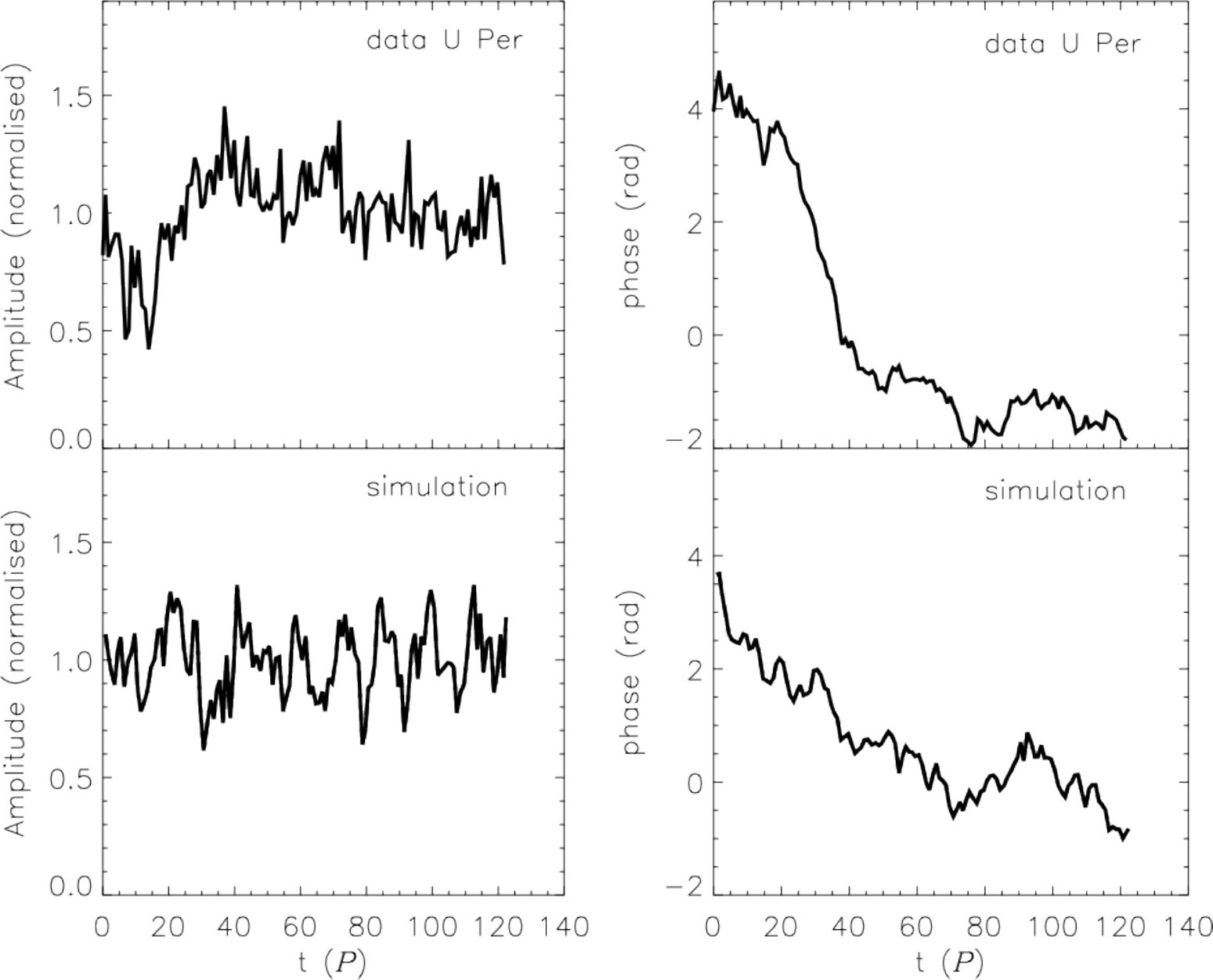}}
	\end{minipage}
	\caption{Top: amplitude (left) and phase (right) for U~Per as a function of time measured in units of the pulsation period, $P=1/f_0$, where $f_0=0.003142$~c/d was taken to be the frequency at which the power is maximum. The amplitude has been normalised by the average amplitude in the period considered. Bottom: the same as the top panels but for one realisation of the simulation performed for U~Per (see text for details).
	}
	\label{fig:amp_phase}
\end{figure}

In order to test the above proposition, we first derive from the data the rms of the relative amplitude variation, $\sigma_{\Delta A/ A}$, and the rms of the phase variation, $\sigma_{\Delta\varphi}$ after the total observation time $t_{\rm obs}\approx 107$~yr. 
To compute the amplitude and phase time variations we subdivided the light curve in sections of roughly one pulsation period ($\approx 320$~d) and performed sine-wave least-square fits to each subsection using the ``Amplitude/Phase variations" functionality of Period04. The retrieved amplitude, normalised by the average value, $\langle A\rangle=1.45$~mag, is shown on the top left panel of Fig.~\ref{fig:amp_phase} and the corresponding phase measurements in the top right panel of the same figure. 

Since the phase variation follows a random walk, $\sigma_{\Delta\varphi}$ can be estimated from
\begin{equation}
\sigma_{\Delta\varphi}\sim\sqrt{N}\times\sqrt{\langle(\Delta\varphi_i)^2\rangle}=\sqrt{\sum_{i=1}^{N}{(\Delta\varphi_i)^2}},
\label{sigmaphi_c}
\end{equation}
where $\Delta\varphi_i$ is the phase difference computed between two consecutive measurements, $i$ and $i-1$, $N$ is the total number of phase difference measurements, and the brackets $\langle\rangle$ indicate an average over all values of $\Delta\varphi$. Unlike the phase, the amplitude loses memory of past perturbations after a time $\sim\eta^{-1}$, varying always close to the mean. As a result, $\sigma_{\Delta A/A}$ can be estimated from 
\begin{equation}
\sigma_{\Delta A/A}\sim\sqrt{\sigma^2_{\Delta A_i/A}}=\sqrt{\sigma^2_{A_i}/\langle A \rangle},
\label{sigmaa_c}
\end{equation}
where $\Delta A_i$ is the difference between the i$^{th}$ measurement of the amplitude and the amplitude average $\langle A\rangle$.

From the amplitude and phase shown in Fig.~\ref{fig:amp_phase} and Eqs~(\ref{sigmaphi_c})-(\ref{sigmaa_c}), we derive $\sigma_{\Delta A/A}=0.18$ and $\sigma_{\Delta\varphi}=2.1$ for U~Per. To verify the sensitivity of these results to the size of the light-curve sections considered in the fits, we repeat the calculation dividing the light curve in sections of 2, 3, and 4 pulsation periods, at a time. Moreover, in the case of fits to light-curve sections extending for one pulsation period, we also determine $\sigma_{\Delta A/A}$ and $\sigma_{\Delta\varphi}$ from phase and amplitude variations computed from non-consecutive measurements (specifically, taking values distanced by 2, 3, and 4 measurements). Changing the extent of the light-curve sections or the distance between amplitude and phase measurements within the limits above results in values of  $\sigma_{\Delta A/A}$ varying between 0.14 and 0.18 and of $\sigma_{\Delta\varphi}$ varying between 1.9 and 2.3.

According to our simulations, which we shall detail below, the values of $\sigma_{\Delta A/A}$ above imply that the star is near the classical limit discussed in Sec.~\ref{class}. Thus, the half width at half maximum of the Lorentzian is $\sim\Gamma_c$, which can be estimated from $\sigma_{\Delta\varphi}$ through Eq.~(\ref{gammac}). 
Doing so, we find $\Gamma_c=8.96\times10^{-6}$~d$^{-1}$ for the phase variability shown in Fig.~\ref{fig:amp_phase}. Given that the resolution of the power spectral density is $1/(2458909.522-2419771.74)\approx 2.6\times10^{-5}$~d$^{-1}$, the Lorentzian is not expected to be fully resolved for U~Per. In fact, if we assume that the transition from an unresolved to a resolved Lorentzian occurs when the total observing time reaches $T_{\rm res}=1/\pi\Gamma$ \cite[e.g.][]{basuandchaplin17}, our estimate of the Lorentzian's half width implies $T_{\rm res}=35526$~d. For the Lorentzian to be fully resolved, the observing time would need to be about one order of magnitude longer than $T_{\rm res}$, but in the case of U~Per the observations run for 39138~d, so for a time only about $10\%$ longer than the estimated $T_{\rm res}$. Unfortunately, this seems to be also the case for the other Mira stars with long-term observations available in the AAVSO database. Thus, it seems that for the pulsation periods typical of Mira stars and the level of stochasticity observed, the many decades-long observations available are not enough to fully resolve the Lorentzian in the power spectral density, despite the stochastic impact being observed in the pulsation amplitude and phase of these stars. 

While the Lorentzian profile of U~Per is not properly resolved, there is no limitation to the length of simulated observations. Hence, we can simulate a star with levels of amplitude and phase variability similar to those inferred for U~Per and verify if the power spectral density from those simulations is consistent with the one derived from the light curve of U~Per, once the simulated light curve is limited to the true length of the observations. To that end, we simulate 5000 pulsation cycles of a star with the pulsation frequency observed in U~Per, $A_N/A=0.0595$, and $\eta=0.695\,P^{-1}$. With this input the theoretical rms of the relative amplitude and phase variation of the simulation are $\sigma^s_{\Delta A/A}=0.18$ and $\sigma^s_{\Delta\varphi}=2.3$, respectively.
We then cut the simulation is sections of 123 pulsation cycles, each corresponding to one realisation with the length of the observations, and interpolate each section on the observation times, to guarantee that the effect of uneven time sampling and gaps is incorporated into the simulations. The amplitude and phase variability differ from realisation to realisation, as expected for a stochastic process. We show an example of that variability for one realisation that resembles  the observations of U~Per in the lower panels of Fig.~\ref{fig:amp_phase}. The values of $\sigma_{\Delta A/A}$ and $\sigma_{\Delta\varphi}$ for this realisation computed in exactly the same manner as done for the observations are $\sigma_{\Delta A/A}=0.15$ and $\sigma^s_{\Delta\varphi}=2.1$. 

\begin{figure} 
	\begin{minipage}{1.\linewidth}  
		\rotatebox{0}{\includegraphics[width=1.0\linewidth]{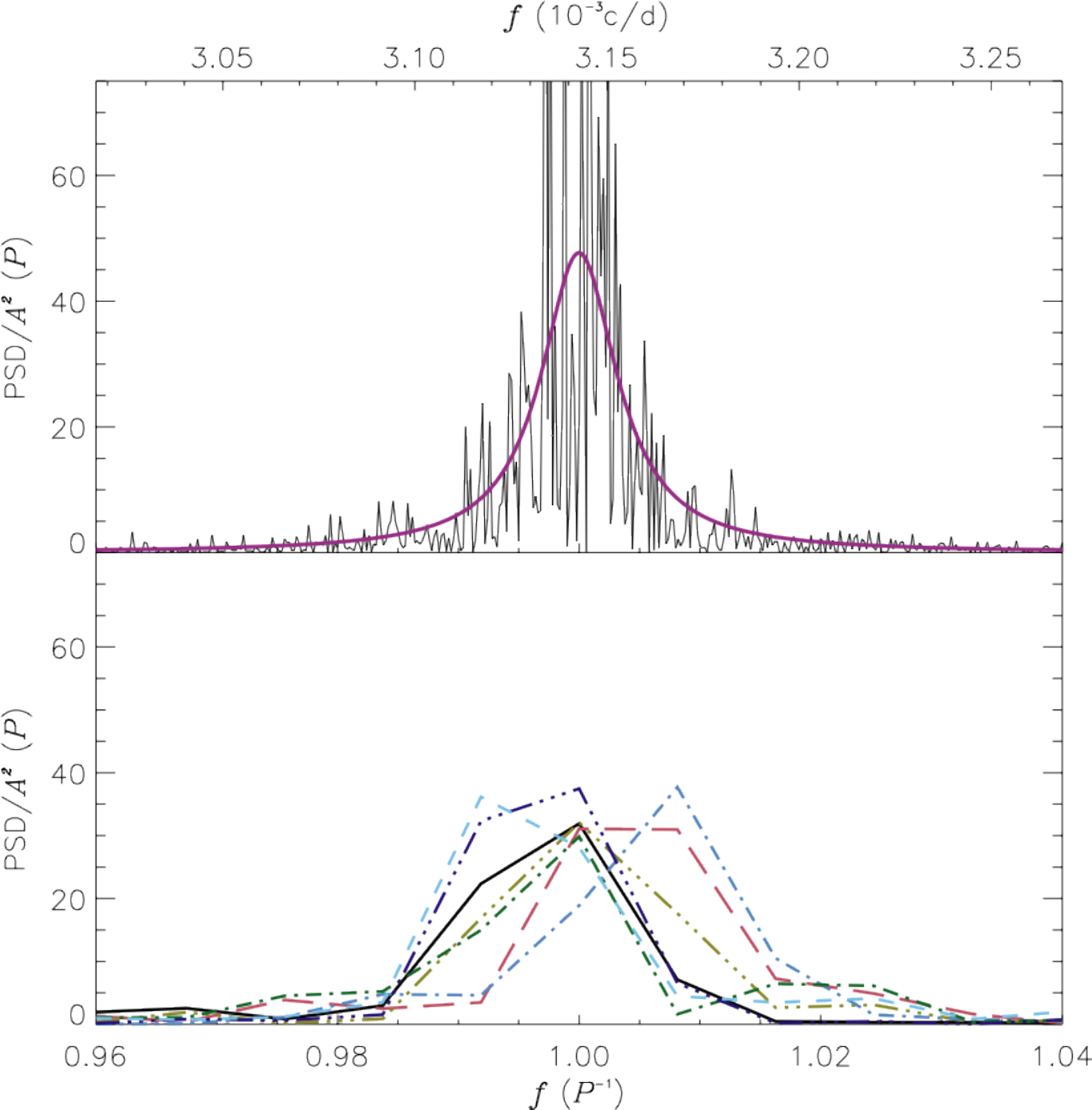}}
	\end{minipage}
	\caption{Amplitude-normalised power spectral density (PSD/$A^2$) in units of the pulsation period. The lower horizontal axis shows the frequency in units of the inverse of the pulsation period and the top horizontal axis in $10^{-3}$~cycles per day. Top panel: PSD/$A^2$ for the full simulation of 5000 pulsation cycles (grey) and the Lorentzian described by Eq.~(\ref{lor_clas}) with the parameters adopted for the simulation (pink):$f_0=0.003142$~c/d (=1$P^{-1}$), $\Gamma_{\rm c}=1.1\times 10^{-5}$~d$^{-1}$ (=0.036$P^{-1}$). Lower panel: comparison between the observed PSD/$A^2$ (black, continuous, thick line) and the PSD/$A^2$ for six different realisations of the simulation, all 123 pulsation cycles long (thick coloured lines, different line styles). In the case of the observed data, the time-averaged amplitude was used as an estimate of $A$ in the normalisation. The observed power spectral density is hardly distinguishable from the simulation representative set shown. }
	\label{fig:psdsim}
\end{figure}

The amplitude-normalised power spectral density for the full simulation (PSD/$A^2$) is shown in the upper panel of Fig.~\ref{fig:psdsim}, along with the Lorentzian function computed with the parameters of the simulation. 
 If U~Per had been observed long enough for the Lorentzian to be fully resolved, one would expect its PSD to look somewhat similar to what is shown in this simulation. Next we inspect how the amplitude-normalised PSD derived from the observations of U~Per compares with those obtained from the 123 pulsation cycle cuts from the simulation. The results are shown in the lower panel of Fig.~\ref{fig:psdsim}. The thick black continuous line displays the observational results, while the other thick lines, of various line styles and colours, show the results obtained for 6 different realisations of the simulations (all with a length of 123 pulsation cycles).  The different realisations show that the PSD of the simulated data can take a variety of shapes when cuts of a length of 123 pulsation cycles are considered. This is not surprising given the stochastic nature of the process and the fact that the resolution in frequency limits the sampling of the true power density to a few points. It is also evident that the observed PSD is consistent with what the simulations predict. Unfortunately, without a fully resolved Lorentzian profile, the PSD is of little use to estimate the parameters associated with the stochastic impact on the pulsations.  It thus appears that for Mira stars the best procedure to infer the impact of the stochasticity on the pulsations consists in computing the rms of the relative amplitude and phase variations from the light curve and using them to constrain the parameters $A_N/A$ and $\eta$ from Eqs~(\ref{siga_lim}) and (\ref{sigphi}). In addition, a direct comparison with model simulations is a fruitful approach to test the robustness of the method used to infer the rms of the relative pulsation amplitude variation and of the phase variation.


\subsection{The semiregular variable L2 Pup}
\label{l2pup}
\begin{figure} 
	\begin{minipage}{1.\linewidth}  
		\rotatebox{0}{\includegraphics[width=1.0\linewidth]{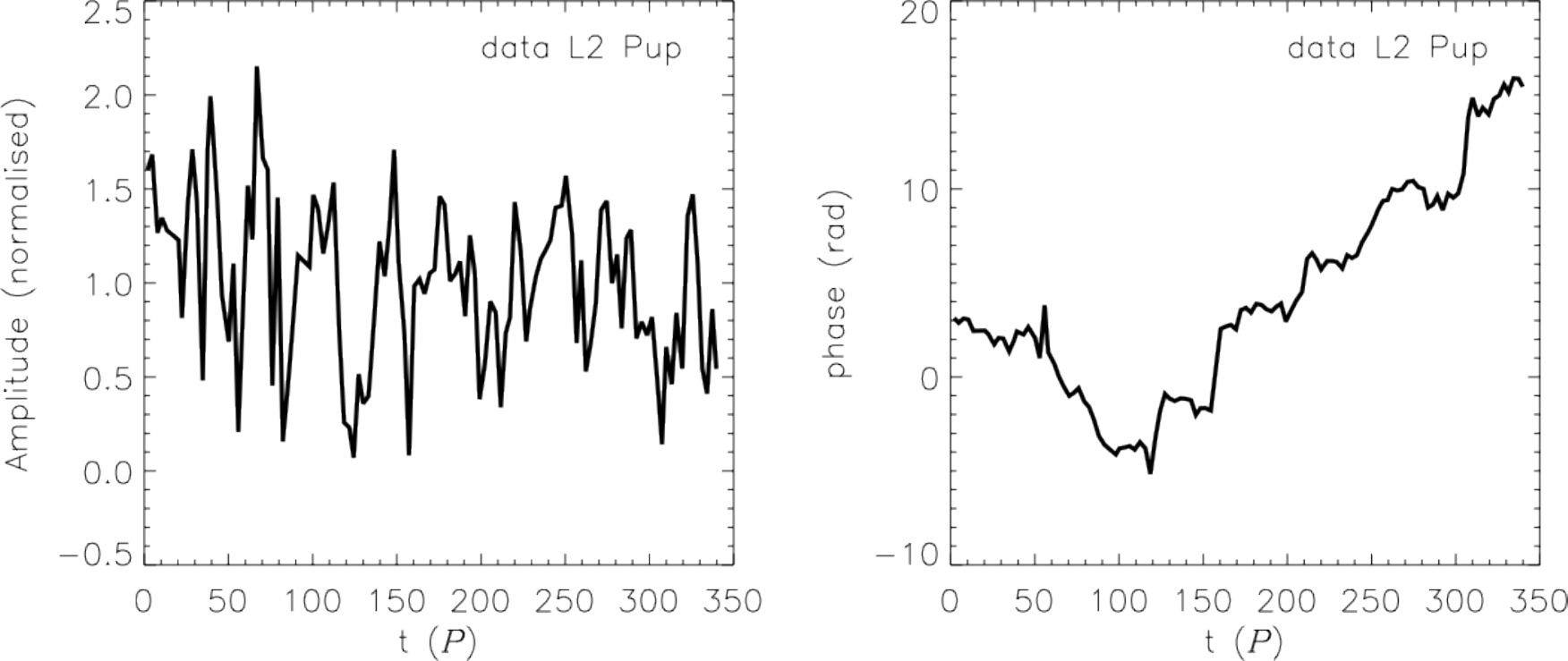}}
	\end{minipage}
	\caption{Amplitude (left) and phase (right) for L2~Pup as a function of time measured in units of the pulsation period, $P=1/f_0$, where $f_0=0.0072$~c/d. The amplitude has been normalised by the average amplitude in the time interval considered. 
	}
	\label{fig:amp_phase_l2pup}
\end{figure}

The data considered for L2~Pup extends for nearly 129 years (from JD 2411839.4  to JD 2458870.074), covering about 342 pulsation cycles. {Significant dimming episodes, presumably due to extinction by circumstellar dust \citep{bedding02}, can be seen in the light curve. The long time scales associated with this phenomenon, which is external to the star, allow us to easily separate its signature from that of the pulsations in the power spectrum.}  The amplitude and phase derived from the data, following the same procedure as for U~Per are shown in Fig.~\ref{fig:amp_phase_l2pup}, where, as before, the amplitude has been normalised by its average value, $\langle A\rangle=0.5$~mag. For the case shown in the figure we have subdivided and fitted light curve sections of roughly 3 pulsation cycles ($\approx$~417~d). We did not consider fits to light curve sections of one pulsation cycle, as we did for U~Per, because gaps of length comparable to half a pulsation cycle are common in the data of L2~Pup. The rms variance of the relative amplitude variation computed from Eq.~(\ref{sigmaa_c}) for the case shown in Fig.~\ref{fig:amp_phase_l2pup} is $\sigma_{\Delta A/A}=0.4$. From this value we may expect the star to be a good example of the intermediate regime, where both $\Gamma_c$ and $\Gamma_s$ should contribute non-negligibly to the Lorentzian half width at half maximum power, $\Gamma$. 

\begin{figure} 
	\begin{minipage}{1.\linewidth}  
		\rotatebox{0}{\includegraphics[width=1.0\linewidth]{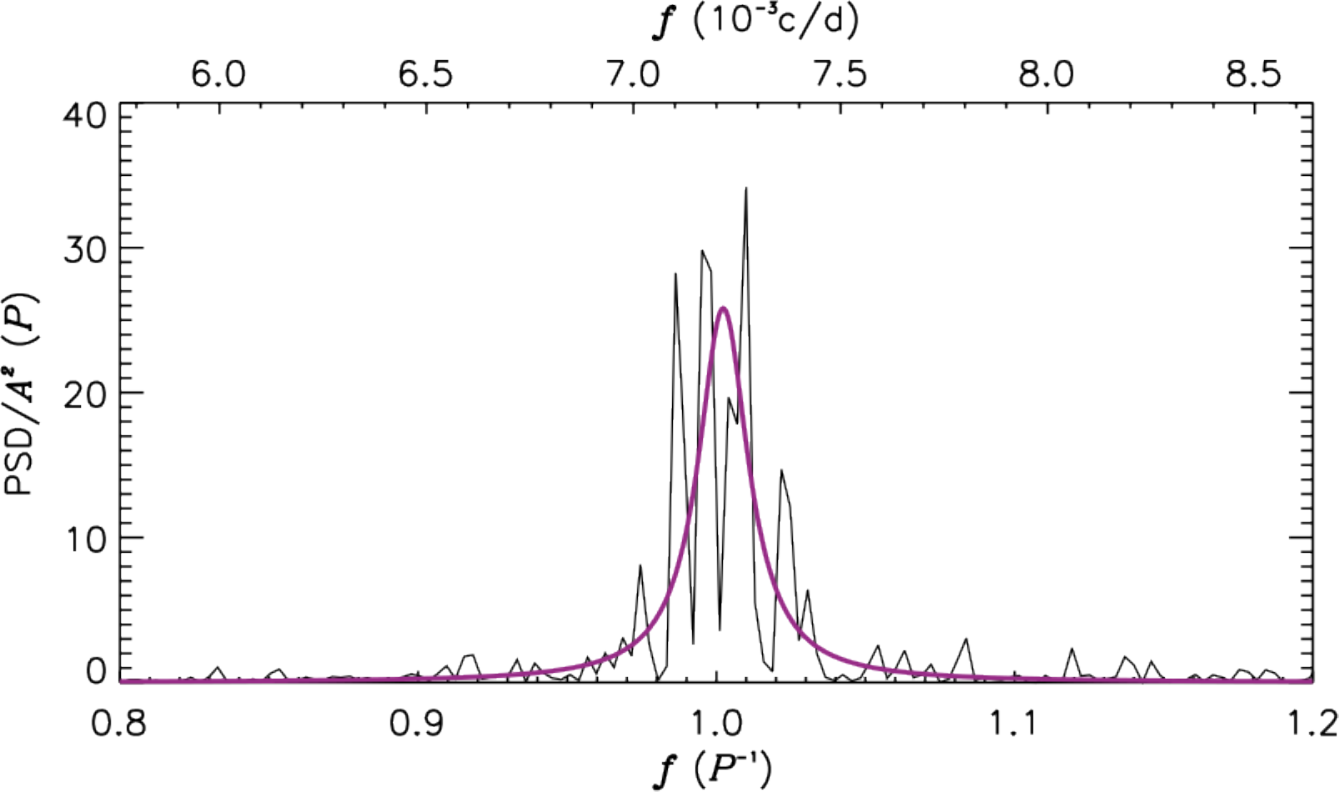}}
	\end{minipage}
	\caption{Amplitude-normalised power spectral density (PSD/$A^2$; black) in units of the pulsation period obtained from the full light curve of L2~Pup (about 342 pulsation cycles). The time-averaged amplitude was used as an estimate of $A$ in the normalisation. The lower horizontal axis shows the frequency in units of the inverse of the pulsation period and the top horizontal axis in $10^{-3}$~cycles per day. Also shown is the Lorentzian function (pink) derived from the fit to the amplitude-normalised power spectral density in the frequency interval shown ([0.8,1.2]$P^{-1}$). 
	}
	\label{fig:psd_l2pup}
\end{figure}

The amplitude-normalised power spectral density for L2~Pup is shown in Fig.~\ref{fig:psd_l2pup}, along with the Lorentzian function that best fits the data. {The fit was performed assuming that the power at a given frequency is described by a $\chi^2$ distribution with two degrees of freedom as discussed by\citet{anderson90}, where we minimised the expression given in their equation (11) considering the model to be a Lorentzian function plus a constant background.} From the fit we find $\Gamma=7.2\times 10^{-5}$~d$^{-1}$ ($=0.01~P^{-1}$). This value is comparable, but slightly smaller than the value of $9.0\times 10^{-5}$~d$^{-1}$ reported by \cite{bedding05}. Repeating the fit to the power spectral density derived from the observation period considered in their work we find  $\Gamma=6.9\times 10^{-5}$~d$^{-1}$, so the difference is not explained by the extent of the data set considered. The explanation for the difference found may stand, instead, on the fact that the data used in \cite{bedding05} was from a different source.  

In the case of L2~Pup, the width of the Lorentzian is $2\Gamma\sim 7/t_{\rm obs}$, or $t_{\rm obs}\sim 11 T_{\rm res}$. Thus, unlike the case of U~Per, the Lorentzian is resolved and we can apply Eqs~(\ref{total_gamma})-(\ref{intermediate}), using $\langle A\rangle$ as an estimate of $A$, to estimate $\Gamma_c=9.8\times 10^{-5}$~d$^{-1}$ and $\Gamma_s=3\times 10^{-4}$~d$^{-1}$. While these values set the star closer to the classical limit than to the stochastic limit, $\Gamma_c$ is not much smaller than $\Gamma_s$ and, thus, the terms involving $\Gamma_s$ cannot be neglected in Eq.~(\ref{intermediate}). From these values, we can further estimate the model parameters $A_N/A=0.12$ and $\eta=0.0019$d$^{-1}$, from which we estimate a mode lifetime, $\tau=\eta^{-1}$, of 1.4~years. This is in contrast with the value that would be derived if the star was assumed to be in the stochastic limit. In that case, $\Gamma_s$ would have been identical to $\Gamma$ and the mode lifetime would have been estimated to be about 6~years, instead of the 1.4 years inferred from our model. {
This implies that the mode lifetimes and quality factors (proportional to the ratio of the mode lifetime to the pulsation period) derived assuming that semiregular variables are in the stochastic limit \citep[e.g.][]{bedding05,stello06}  are likely overestimated and need to be revised.}

\section{Discussion}
\label{discussion}

{The model presented in this work, along with the results from the analysis of L2~Pup,
can be used to interpret observational trends reported in the literature concerning both pulsation amplitudes and mode line widths. In this section we discuss two such trends in the light of our model, namely, the observed pulsation amplitude-period relation and the mode line width-temperature scaling.}

\subsection{Transition between regimes: amplitude behaviour}
{An interesting question to address is where to seek evidence for a transition between the predominantly stochastic to predominantly coherent regimes. The transition occurs where $\Gamma_{\rm c}/\Gamma_{\rm s}=1$, but determining this ratio from stellar data requires an estimate of the underlying coherent amplitude, which may be difficult for stars approaching the transition, as the amplitude variations become very large and stars tend to become multi-mode pulsators. }

{A possible alternative to search for the transition between the predominantly stochastic and the predominantly coherent regimes is to consider the behaviour of the pulsation amplitudes measured from the power spectrum. Inspecting the amplitude of the Lorentzian function in Eq.~\eqref{intermediate} ($A^2/2\pi\Gamma\times (1+2\Gamma_{\rm c}/\Gamma_{\rm s})$), we see that as the transition is approached from the stochastic side, two additional terms are added to what would be the amplitude in the stochastic limit. The second of this terms, proportional to $A^4/A_N^2$, will dominate after the transition, as the classical limit is approached, but at some point before the transition (\itie at shorter periods) a change in the behaviour of the amplitude is expected, as the impact of the coherent driving starts to be noticeable.  Indeed, several studies have noted a change in behaviour in the pulsation amplitude-period relation \citep{banyai13,mosser13,yu20} at a pulsation period ranging from about 5 to 10~d. We conjecture that the change in behaviour reported in these works results from the increasing impact of the coherent driving on the oscillations. We can thus use these observations to estimate the pulsation period at which the transition between the two regimes takes place. }

{According to our model, at $\Gamma_{\rm c}/\Gamma_{\rm s}=1$ the amplitude of the Lorentzian is predicted to be three times larger than what would be expected in the stochastic limit (\itcf Eqs~\eqref{Px} and \eqref{intermediate}). However, as we have seen in Sec. \ref{test}, the Lorentzian amplitude in Eq.~\eqref{intermediate} is somewhat underestimated, so a better estimate of this amplitude ratio can be derived from numerical simulations, from which we find a value close to 4. To estimate the pulsation period at which  $\Gamma_{\rm c}/\Gamma_{\rm s}=1$ we consider the pulsation amplitude-period relation published by \cite{yu20} (their figure 2(a)). Notice that the amplitudes reported in that work were not derived from fitting a Lorentzian function to the power spectral density, as in our model. Nevertheless, we are only interested in amplitude ratios, so we can use these results as a proxy to our test.  The authors fitted a piecewise linear model to the data and found a change in the slope of the pulsation amplitude-period relation at around 4.5~d.  We extrapolate the linear relation they found on the short-period side to derive the pulsation amplitude that would be expected at larger pulsation periods if the pulsations were driven only by the stochastic noise.
Doing so, we find that the ratio between the measured and (extrapolated) stochastic amplitudes is 4 for a pulsation period of about 60~d. Thus, under the assumption that the change in the slope of the pulsation amplitude-period relation reported earlier by several authors results from the additional coherent driving, we conclude that the latter starts to have an impact on the amplitudes at around a period of 5 to 10~d, while the driving is still stochastically dominated, the actual transition between the two driving regimes  ($\Gamma_{\rm c}/\Gamma_{\rm s}=1$ in our model) taking place at a pulsation period of about 60~d.}


\subsection{Mode line width scaling relation}
\begin{figure} 
	\begin{minipage}{1.\linewidth}  
		\rotatebox{0}{\includegraphics[width=1.0\linewidth]{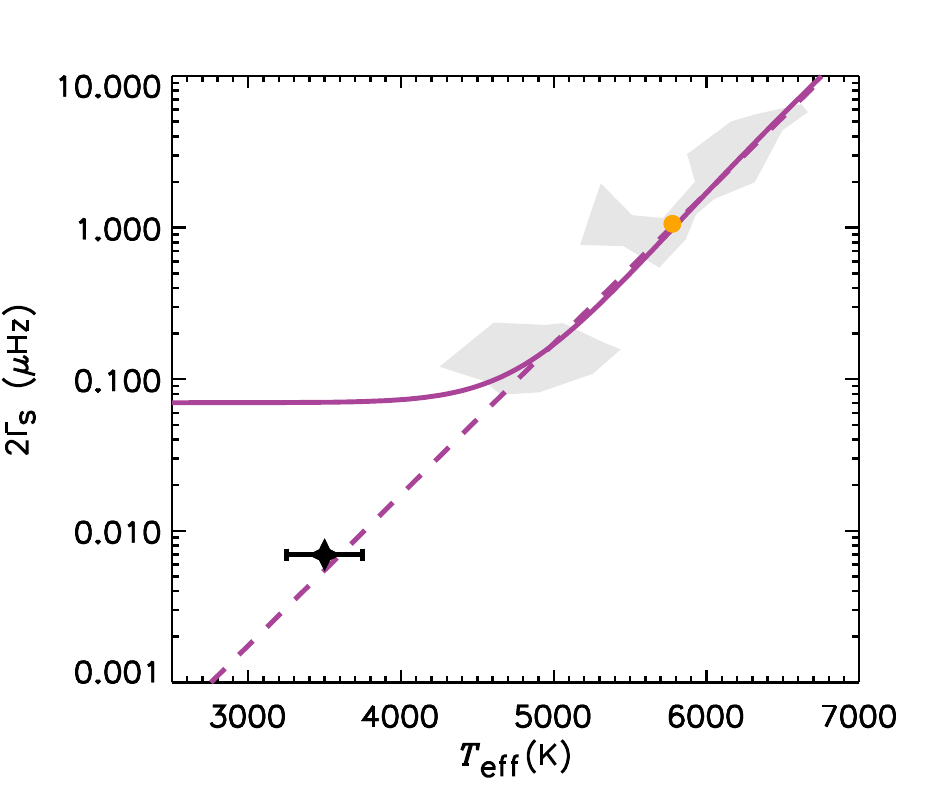}}
	\end{minipage}
	\caption{Comparison between the stochastic component of the line width (2$\Gamma_{\rm s}$) estimated for L2~Pup (black star) and the fits to the line widths at frequency of maximum power derived from solar-like stars at different (earlier) evolutionary status. The two pink curves show the fits performed by \citet{lund17} of a power law function and an exponential function, respectively, to the line widths of stars spanning from the main sequence to the red-giant phase (occupying roughly the position in the diagram marked by the shaded light-grey areas). {The orange circle marks the location of sun \citep{houdek19}.} The value of $\Gamma_{\rm s}$ inferred for L2~Pup is consistent with a single exponential dependence of $\Gamma_{\rm s}$ on effective temperature. 	}
	\label{fig:gammas_all}
\end{figure}

The mode line widths in stochastically-driven pulsators  ($=2\Gamma_{\rm s}$ in this work) have been shown to scale with effective temperature \citep[e.g.][]{chaplin09,baudin11,appourchaux12,corsaro12,lund17}. The inference of $\Gamma_{\rm s}$ for a sample of long period variables, based on the model proposed in this work, will enable this scaling relation to be tested up to the luminous asymptotic giant branch stars, thus, down to temperatures more than 1000~K lower than considered in previous studies. Since we have derived $\Gamma_{\rm s}$ for one star only, we defer to future work a thorough test of the scaling relation. Still, we can verify where L2~Pup stands in comparison to scaling relations published in the literature. In order to do so, we adopt the effective temperature of $T_{\rm eff}=3500\pm 250$~K published by \citet{kervella14,kervella16} for L2~Pup, which the authors derived from the star's near infrared $H$ and $K$ band magnitudes and the angular diameter determined from interferometry. 
Figure~\ref{fig:gammas_all} shows the position of L2~Pup (black star) on the $2\Gamma_{\rm s}-T_{\rm eff}$ diagram, along with the position of the sun (orange circle). Two scaling relations published by \citet{lund17} are also shown, one based on a power law (pink continuous line)  and the other on an exponential function (pink dashed line) fitted to {\it Kepler} data. To derive these fits the authors used line widths and effective temperatures from the {\it Kepler} Legacy sample, consisting of main-sequence and subgiant stars, and from {\it Kepler} red giants analysed by  \citet{corsaro12} and \citet{Handberg17}. The result for L2~Pup does not support the power law fit with the parameters derived based on the data considered by \citet{lund17}, but confirms that a single exponential function represents well the dependence of $\Gamma_{\rm s}$ on the effective temperature, down to temperatures as low as 3500~K.

\section{conclusions}
\label{conclusions}
This work provides a {unifying} description of the limit power spectral density for pulsations in the presence of stochastic noise, from main-sequence solar-like pulsators to the luminous AGB Mira stars. The model, first proposed by  \cite{avelino20} {(Paper~I)}, is based on an internally driven damped harmonic oscillator and predicts that in all cases, the limit power spectral density of a pulsation mode in the presence of stochastic noise is described by a Lorentzian function. Generally, the half width at half maximum power of the Lorentzian function, $\Gamma$, depends both on the damping constant (thus, on mode lifetime) and on the amplitude of the stochastic noise. Nevertheless, as the stochastic limit is approached, $\Gamma$ tends to $\Gamma_{\rm s}=\eta/(2\pi)$, thus becoming independent of the amplitude of the stochastic noise, and as the classical limit is approached, $\Gamma$ tends to $\Gamma_{\rm c}=(A_N/A)^2\times\omega_0/(2\pi)$, thus, becoming independent of the mode lifetime. 

The model allows us to introduce a quantitative classification of pulsating stars in the presence of stochastic noise. When $\Gamma_{\rm c}/\Gamma_{\rm s} > 1$, the driving is predominantly stochastic and the stars are classified as solar-like pulsators, while when $\Gamma_{\rm c}/\Gamma_{\rm s} < 1$, the driving is predominantly coherent and the stars are classified as classical pulsators. In this context, when no coherent driving is present, the star is a pure solar-like pulsator and when there is no source of stochastic noise, the star is a pure classical pulsator. We note, however, that the latter case was not considered in this study, as we have always assumed the presence of stochastic noise. 

The model has been applied in the analysis of the observations of the Mira star U~Per and the semiregular star L2~Pup. In the case of the Mira star, a classical pulsator according to our classification scheme, we have shown that the Lorentzian is not fully resolved. Unfortunately, that seems to be the case also for other Mira stars. Nevertheless, the model parameters for U~Per could still be estimated through inspection of the amplitude and phase variability, and used to perform model numerical simulations. Comparison of the simulation results with the power spectral density for U~Per corroborates our model predictions. In the case of L2~Pup, the Lorentzian function is resolved and we could infer both $\Gamma_{\rm c}$ and $\Gamma_{\rm s}$ from the inspection of the power spectral density and the amplitude variability. We could then derive the mode lifetime and show that it is significantly shorter than what would have been derived if we had wrongly assumed that the width of the Lorentzian depended only on $\eta$, as in the case of solar-like pulsators. In fact, we have confirmed that in this star $\Gamma_{\rm c}/\Gamma_{\rm s} < 1$, thus the driving is predominantly coherent. Hence, according to our classification, L2~Pup is a classical pulsator. 

{Our model also provides a natural explanation for the change in behaviour of the observed pulsation amplitude-period relation reported in several earlier works. Based on our model, we argue that the change in the amplitude behaviour seen at periods around 5 to 10~d is associated to the beginning of a non-negligible contribution of the coherent driving which increases towards longer periods with the actual change between the stochastically- and coherently-dominated regimes occurring around pulsation periods of 60~d.} Moreover, with the results inferred for L2~Pup we are able to extend the test to the scaling relation between the mode line width and the effective temperature, confirming that a single exponential relation provides a good representation of the scaling relation down to temperatures about 1000~K cooler than previously considered in the literature. 

Finally, we emphasise that our model can be considered in the study of any stellar pulsator holding a source of stochastic noise, independently of how important stochasticity is to the driving of the pulsations. Moreover, given our findings, we note that stochastic noise should be added to the list of possible sources for the pulsation phase variability observed in classical pulsators across the HR diagram, including those in the main-sequence and in later stages of evolution. 

\section*{Data availability}
The data underlying this article will be shared on reasonable request to the corresponding author.

\section*{Acknowledgements}
We acknowledge with thanks the variable star observations from the AAVSO International Database contributed by observers worldwide and used in this research. P. P. A. thanks the support from FCT -- Funda{\c c}\~ao para a Ci\^encia e a Tecnologia -- through the Sabbatical Grant No. SFRH/BSAB/150322/2019. M. S. Cunha is supported by national funds through FCT in the form of a work contract. This work was supported by FCT through the research grants UIDB/04434/2020, UIDP/04434/2020 and PTDC/FIS-AST/30389/2017, and by FEDER - Fundo Europeu de Desenvolvimento Regional through COMPETE2020 - Programa Operacional Competitividade e Internacionalização (grant: POCI-01-0145-FEDER-030389). W.~J.~C. acknowledges support from the UK Science and Technology Facilities Council (STFC). Funding for the Stellar Astrophysics Centre
is provided by The Danish National Research Foundation (grant
agreement no. DNRF106).




\bibliography{solar-like.bib} 


\end{document}